\documentclass{aa}  

\usepackage{graphicx}
\usepackage{txfonts}
\usepackage{ulem}
\usepackage{placeins}
\usepackage{amsmath}
\usepackage{booktabs}
\usepackage{natbib}
\usepackage{xcolor}
\usepackage[colorlinks=true,citecolor=blue,linkcolor=magenta,urlcolor=blue]{hyperref}
\newcommand{\teff}{$T_{\rm eff}$}
\newcommand{\met}{[M/H]}

\newcommand{\dc}{$\Delta C$}

\newcommand{\msun}{$\text{M}_\odot$}
\newcommand{\cmapx}{$\Delta_{\rm F275W,F814W}$}

\begin{document} 

   \title{A stellar census in globular clusters with MUSE.\\ Metallicity spread and dispersion among first-population stars\thanks{Based on observations collected at the European Organisation for Astronomical Research in the Southern Hemisphere, Chile (Program IDs 094.D-0142(B), 095.D-0629(A), 096.D-0175(A), 097.D-0295(A), 098.D-0148(A), 099.D-0019(A), 0100.D-0161(A), 0101.D-0268(A), 0102.D-0270(A), 0103.D-0204(A), 0103.D-0545, 0104.D-0257(B), and 105.20CR.002)}}

   \author{M. Latour\inst{1},
           S. Kamann\inst{2},
           S. Martocchia\inst{3},
           T.-O. Husser\inst{1},
           S. Saracino\inst{2},
           \and
           S. Dreizler\inst{1}
          }
     
   \institute{Institut für Astrophysik und Geophysik, Georg-August-Universität Göttingen, Friedrich-Hund-Platz 1, 37077 Göttingen, Germany\\
              \email{marilyn.latour@uni-goettingen.de}
         \and
                  Astrophysics Research Institute, Liverpool John Moores University, 
         IC2 Liverpool Science Park, 146 Brownlow Hill, Liverpool, L3 5RF, United Kingdom
         \and
            Astronomisches Rechen-Institut, Zentrum für Astronomie der Universität Heidelberg, Mönchhofstraße 12-14, D-69120 Heidelberg, Germany \\
             }

   \date{Received -- 9/30/2024; Accepted -- 01/15/2025}

 
  \abstract
   {Multiple populations are ubiquitous in the old massive globular clusters (GCs) of the Milky Way. It is still unclear how they arose during the formation of a GC. The topic of iron and metallicity variations has recently attracted attention with the measurement of iron variations among the primordial population (P1) stars of Galactic GCs.  }
   {We want to explore the relationship between the metallicity of the P1 stars on the red-giant branch (RGB) of Galactic GCs and their $\Delta_{\rm F275W,F814W}$ pseudo-color. We also want to measure the metallicity dispersion of P1 and P2 stars.}
   {We use the spectra of more than 8000 RGB stars in 21 Galactic GCs observed with the integral-field spectrograph MUSE to derive individual stellar metallicities [M/H]. For each cluster, we use the \textit{Hubble Space Telescope} (HST) photometric catalogs to separate the stars into two main populations (P1 and P2).  We measure the metallicity spread within the primordial population of each cluster by combining our metallicity measurements with the stars $\Delta_{\rm F275W,F814W}$ pseudo-color. We also derive metallicity dispersions ($\sigma_{\rm [M/H]}$) for the P1 and P2 stars of each GC.  }
   {In all but three GCs, we measure a significant correlation between the metallicity and the $\Delta_{\rm F275W,F814W}$ pseudo-color of the P1 stars such that stars with larger $\Delta_{\rm F275W,F814W}$ have higher metallicities. We measure metallicity spreads that range from 0.03 to 0.24 dex and correlate with the GC masses. As for the intrinsic metallicity dispersions, when combining the P1 and P2 stars, we measure values ranging from 0.02 dex to 0.08 dex and correlate very well with the GC masses. The two clusters showing the largest $\sigma_{\rm [M/H]}$ are NGC\,6388 and NGC\,6441. We compared the metallicity dispersion among the P1 and P2 stars and found that the P2 stars have metallicity dispersions that are smaller or equal to that of the P1 stars. }
   {We present a homogeneous spectroscopic characterization of the metallicities of the P1 and P2 stars in a set of 21 Galactic GCs. 
We find that both the metallicity spreads of the P1 stars (from the $\Delta_{\rm F275W,F814W}$ spread in the chromosome maps) and the metallicity dispersions ($\sigma_{\rm [M/H]}$) correlate with the GC masses, as predicted by some theoretical self-enrichment models presented in the literature. }


 \authorrunning{Latour et al.}
 \titlerunning{Metallicity spread and dispersion among P1 stars}    

   \maketitle
%

\section{Introduction}\label{sec:intro}

The formation of massive star clusters has been puzzling astrophysicists over the past decades. Despite considerable progress recently gained through observations and hydrodynamical simulations, a wealth of open questions remains. This is probably best illustrated by the enigma of multiple populations, star-to-star abundance variations in light elements (such as C, N, O, Na) that are ubiquitous in massive clusters older than $\sim$2~Gyr, such as the Galactic globular clusters or intermediate-age star clusters in the Magellanic Clouds \citep[see][for recent reviews]{2018ARA&A..56...83B,Gratton2019}. Despite a multitude of observational studies and a number of proposed scenarios, the mechanisms underlying the formation of multiple populations are still unknown.


From an observational perspective, the ubiquity of the multiple populations phenomena has been confirmed both photometrically and spectroscopically \mbox{\citep[e.g.,][]{2009A&A...505..117C}}.
The work of \citet{mil17}, making use of UV and optical magnitudes from the \textit{Hubble Space Telescope} (HST) in the form of chromosome maps (which are pseudo-two-color diagrams), confirmed the presence of multiple populations in a sample of 57 Galactic globular clusters. In all of these clusters, the RGB stars have a spread in color larger than what is expected from photometric errors only. The pseudo-color, (i.e., ($m_{\rm F275W} - m_{\rm F336W}$)$-$($m_{\rm F336W} - m_{\rm F814W}$) ), allows an efficient separation of the two principal population of stars in GCs; the so-called first and second populations (further referred to as P1 and P2). This is because this pseudo-color mainly traces nitrogen variations \citep{mil17}. From matching photometric and spectroscopic properties of RGB stars among these populations, it was found that the P1 stars are those with an atmospheric chemistry showing primordial abundances (i.e., a scaled solar-like abundance pattern, except for some degree of $\alpha$-enhancement) and the P2 stars are those with an "anomalous" abundance pattern, most notably, enhancement in N and Na and depletion in O (see, e.g. \citealt{Marino19}, \citealt{milone2015_2808}, \citealt{Carretta_2024}.) A few GCs have an additional population (P3) that is not only visible on the chromosome map but also on their CMD as a red-RGB which, in some cases, also connects with a fainter subgiant branch (SGB). The most notorious case is $\omega$~Centauri but other GCs such as NGC\,1851 and NGC\,5286 also have a distinct red-RGB and a faint SGB. These clusters were termed as Type II (also anomalous or iron-complex, \citealt{Johnson2015}) GCs and the stars belonging to their red-RGB are typically enriched in elements produced via the slow neutron-capture process ($s$-elements) such as Ba and La, and possibly also in their iron content (see, e.g. \citealt{marino2015} their Table 10), although the latter point is still under debate (see \citealt{Carretta_2023} and references therein).

What came unexpected was the realization that in the majority of the GCs from \citet{mil17}, the P1 stars themselves have a spread in $F275W - F814W$ that is larger than expected from measurement errors. This means that even this "primordial" population is not consistent with a simple stellar population. Spectroscopic analyses of P1 stars found in the chromosome maps were scarce at the time 
and no particular chemical species could be found to explain such a color spread. Based on the photometric properties of the P1 stars in three GCs showing different color spread in their chromosome maps, \citet{Lardo18} found that a spread in He abundance and a small range of N abundances could explain the extended color distribution of the primordial population. An extensive comparison of abundance values from literature with the position of stars in the chromosome maps of 29 GCs performed by \citet{Marino19} found no evidence of light-element variations among the P1 stars and the authors suggested instead that variations in iron or helium could explain the color spread. Dedicated spectroscopic observations of P1 stars in NGC\,2808 also found no evidence of light-element variations \citep{cabrera19}. 
Additional spectroscopic investigations of P1 stars in NGC\,3201 suggested that a small spread in iron (by $\sim$0.1 dex) is present among these stars and could explain their pseudo-color distribution in the chromosome map \citep{Marino19_3201}. Further spectroscopic and photometric investigations supported this hypothesis (e.g., \citealt{Husser20,lardo22}) and the presence of a small iron-spread among the primordial stellar population is now strongly favored as opposed to He-variations (see also \citealt{Tailo19}). 
Iron variations among P1 stars have been measured from high-resolution spectroscopy for a handful of stars in three GCs so far, NGC\,3201, NGC\,2808, and NGC\,104 (47 Tuc). The iron variations were found to be in the range of 0.1 to 0.15 dex \citep{Marino19_3201,Marino23_47tuc,Lardo2023}. Abundances of up to 24 atomic species were also measured in the stars of NGC\,3201 and NGC\,104 and for most species a positive correlation with the pseudo-color of the star was found, suggesting that not only iron varies but the overall stellar metallicity as well.
Finally, by comparing the photometric properties of the P1 stars, essentially the width of their pseudo-color distribution in the chromosome maps, with isochrones of varying metallicity, \citet{Legnardi22} estimated metallicity variations in 55 GCs. They found a wide range of values: from less than 0.05 dex to $\sim$0.30 dex.  

The presence of metallicity (or iron) spreads within P1 stars has implications for our view of GCs as a whole.
Apart from a few particular cases of massive clusters showing a clear iron-spread, and possible age-spread as well, such as $\omega$ Cen, NGC\,6715 (M54), and Terzan 5 (see, e.g. \citealt{Johnson2010,alfaro2019,Ferraro2009,pfeffer2021}), the traditional view is that GCs do not show a significant spread in iron abundances. For example, \citet{carretta2009} established an upper limit of 0.05 dex for possible iron-spread among 19 GCs based on direct iron measurements from high-resolution spectroscopy. The recent catalog of iron dispersion compiled in \citet{Bailin19_cat}, based on selected iron abundances from literature, showed that iron-spreads among GCs are indeed modest ($<$ 0.1 dex) but significantly different than zero in the majority of cases. \citet{yong2013} also measured a small (0.03 dex) but significant metallicity spread (for iron and a dozen additional species) among RGB stars in NGC\,6752 using high-precision differential abundance measurements. 
The measurement of internal iron spread of such small amplitude is notoriously difficult. At times, studies have claimed to find large ($>$0.1 dex) iron spreads in some GCs,
but these were often not corroborated by further investigations. Even the issue of whether the P3 population in some type II GCs is enhanced in iron or not is still a matter of debate \citep{mucciarelli2015_m22,carretta2022_ngc6388,varagas2022_ngc362,McKenzie2022_ngc6656}. 
Artificially large iron spreads can be caused by a few factors, such as the inclusion of AGB stars among a sample of RGB objects \citep{mucciarelli2015}, the method used to determine the surface gravity of the stars \citep{mucciarelli2015_m22}, or even the presence of intrinsic luminosity variations in some stars of the studied sample \citep{albornoz2021}. 

The presence of intrinsic iron variations is of particular interest for the theoretical modeling of GC formation.
It is commonly assumed that massive star clusters form hierarchically, via the merging of smaller sub-clusters \mbox{\citep[e.g., see][and references therein]{2014CQGra..31x4006K}}. In this scenario, iron variations within clusters could point to chemical inhomogeneities of the interstellar medium within the spatial scales that form massive clusters. Alternatively, it has been suggested that clusters self-enrich, for example via core-collapse supernovae (SNe) of the first and most massive stars formed in the cluster while the least massive stars are still in their formation process \citep{Morgan1989,Wirth2021}.
In line with the hierarchical scenario of cluster formation, it has been argued that sub-clusters produce various levels of iron enrichment before merging to form the final GC \mbox{\citep{bailin2018,Mckenzie21}}.
Such modeling of GC formation can reproduce, qualitatively, the mass-metallicity relationship observed among the massive metal-poor ([Fe/H]$\sim-$1.5) GCs around galaxies (see e.g., \citealt{Bailin2009,Strader2008}; but see \mbox{\citealt{2018MNRAS.480.3279U}} for an alternative explanation).
However, the extent to which even massive star clusters can self-enrich is still debated. Star formation is expected to be suppressed by stellar winds even before the onset of SNe \mbox{\citep{2021MNRAS.506.3882S}} and indeed, young massive clusters in the local Universe are found to be gas free \mbox{\citep{2014ApJ...795..156W,2015MNRAS.448.2224C}}. Recent hydrodynamical simulations \mbox{\citep[e.g.,][]{lahen2024}} find that while enrichment in light elements through stellar wind material from the short-lived massive stars appears feasible, the clusters do not significantly enrich in iron or other heavy elements via SNe.

In this work, we make use of our MUSE spectral database of GC stars to investigate the topic of metallicity dispersion and variations within first-population stars. We select P1 stars from the red-giant branch of 21 GCs based on their position in the chromosome maps built from HST photometric catalogs \citep{piotto15,nardiello18} and we measure the metallicity of the selected RGB stars from the MUSE spectra. We want to make a clear distinction here between metallicity and iron abundance as these terms are sometimes used interchangeably in the literature. What is measured from our MUSE spectra is an overall solar-scaled metallicity\footnote{except for the $\alpha$-enhancement factor.} and not an iron abundance from individual iron lines. Thus we always refer to it as metallicity ([M/H]) and we keep the use of iron abundance ([Fe/H]) for direct measurements from iron lines, unless specified otherwise.
The paper is organized as follows,
in Sect.~\ref{sec:obs} we describe our spectroscopic data and explain the different aspects of our methodology in Sect.~\ref{sec:met}. Our results on the metallicity spread among the P1 stars and on the metallicity dispersions are presented in Sect.~\ref{sec:res} and discussed in Sect.~\ref{sec:diss}. We briefly conclude in Sect.~\ref{sec:concl}.

\begin{figure*}
\resizebox{\hsize}{!}{
   \includegraphics{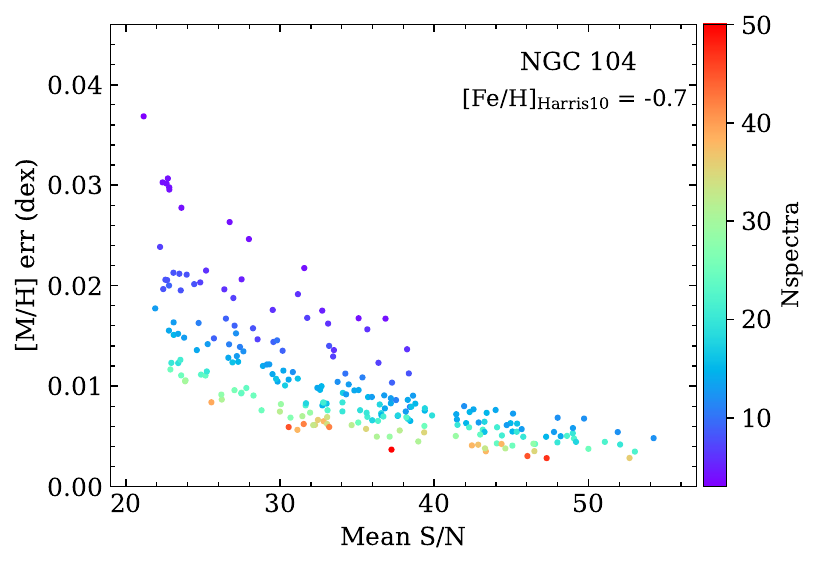}
   \includegraphics{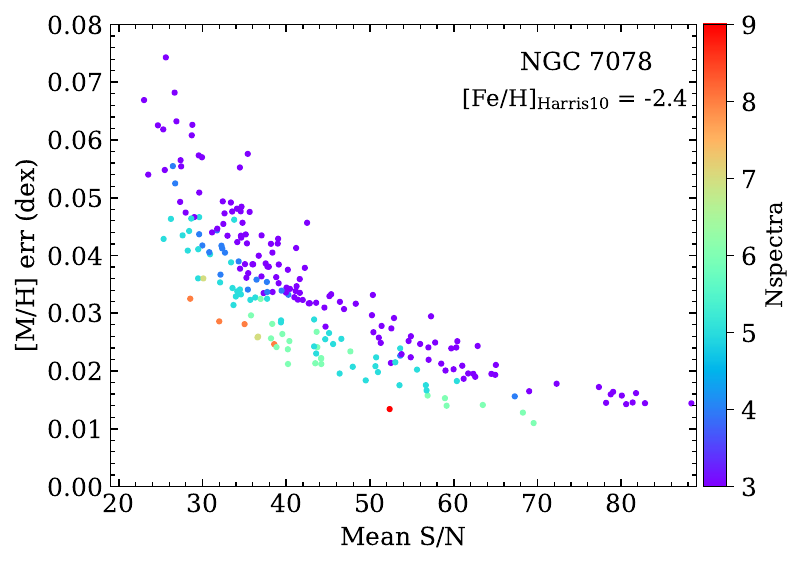}}
     \caption{Mean $S/N$ versus the uncertainty on the metallicity (\met\ err) for a subsample of RGB stars in NGC\,104 and NGC\,7078, color-coded with the number of spectra per star.
     }
     \label{fig:Met_SNR_err}
\end{figure*}
\section{Observational data}\label{sec:obs}

Our target stars have been observed with MUSE as part of the GTO program dedicated to Globular Clusters (PI: S. Dreizler, S. Kamann). Out of the 25 Galactic GCs observed for the survey, 20 of them (listed in Table~\ref{table_results})
have the photometric data necessary to build their chromosome maps and a sufficient number of RGB spectra to perform our analysis. In addition to the GCs observed with the GTO program, we also used the MUSE observations of NGC\,6362 taken as part of GO time (Prop ID: 0103.D-0545, PI: Dalessandro).
MUSE is an integral field spectrograph mounted on UT4 of the VLT and is in operation since 2014 \citep{bacon10}. It features a wide-field-mode (WFM) with a field of view of 1\arcmin $\times$1\arcmin\ at a sampling of 0.2\arcsec\ per pixel and a narrow-field-mode (NFM) covering a smaller field of view (7.5\arcsec $\times$7.5\arcsec) at a sampling of 0.025\arcsec\ per pixel. Both modes result in a spectral coverage of 4750$-$9350 \AA, with a spectral resolution $R \sim 3000$ although this varies slightly across the wavelength range (see \citealt{husser2016}). The GC GTO program targeted the central region of the clusters, covering approximately up to their half-light radii. Depending on the cluster, a varying number of WFM pointings were used to cover the area of interest.
Observations for the GTO program were taken between 2014 and 2022, most of the observations taken after mid-2017 made use of the adaptive optic (AO) system installed on UT4, whenever the observing conditions allowed it. In addition to the WFM pointings made for each GC, ten GCs have one or more additional NFM (with AO) observations located at their very center (see, e.g., \citealt{goettgens21}). However, because of the small field of view of the NFM, these observations contribute only a very small amount of RGB spectra per cluster. 


The data reduction and spectral extraction processes have been described at length in previous papers using spectra from the Globular Cluster GTO program (e.g. \citealt{kamann2018,kamann2016})
The basic data reduction of the MUSE datacubes is performed using the official MUSE pipeline \citep{Weilbacher20}. 
The extraction of the individual stellar spectra is done with \textsc{pampelmuse}\footnote{\url{https://pampelmuse.readthedocs.io/en/latest/about.html}} \citep{Kamann2013,kamann2018} and relies on a reference source catalog to determine the position of each resolved star in the MUSE data. The photometric catalogs used for the GCs included in our study are from the ACS Survey of Globular Clusters \citep{sarajedini2007,anderson08}.

\begin{figure*}
\resizebox{\hsize}{!}{
   \includegraphics{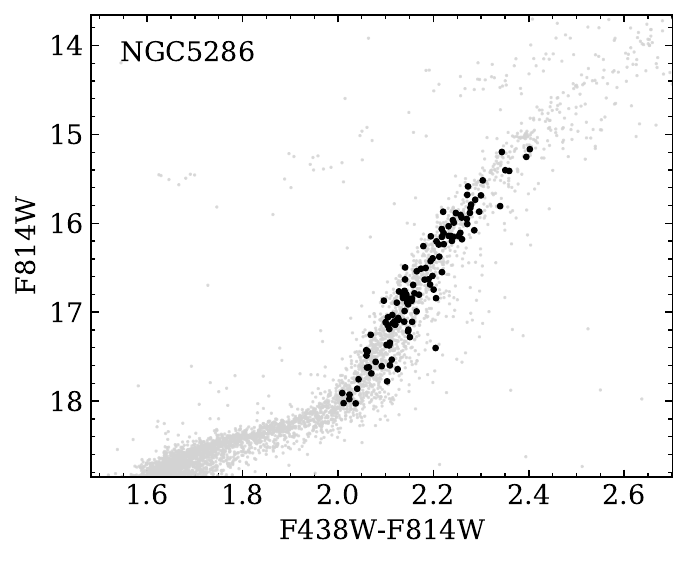}
   \includegraphics{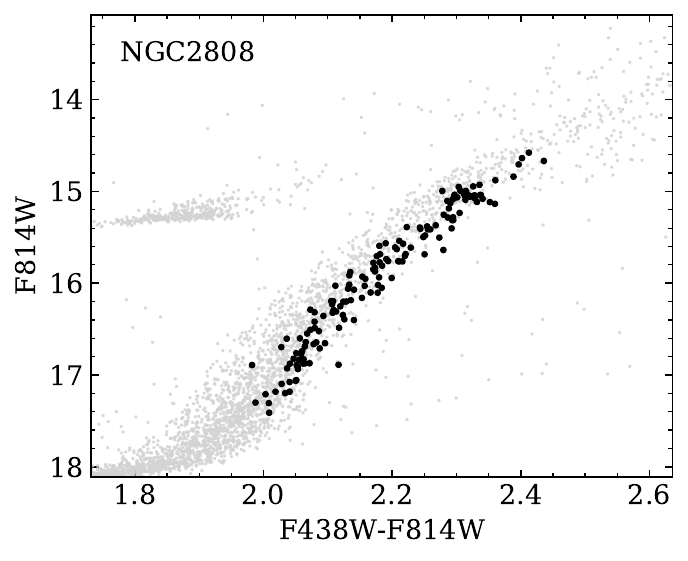}
   \includegraphics{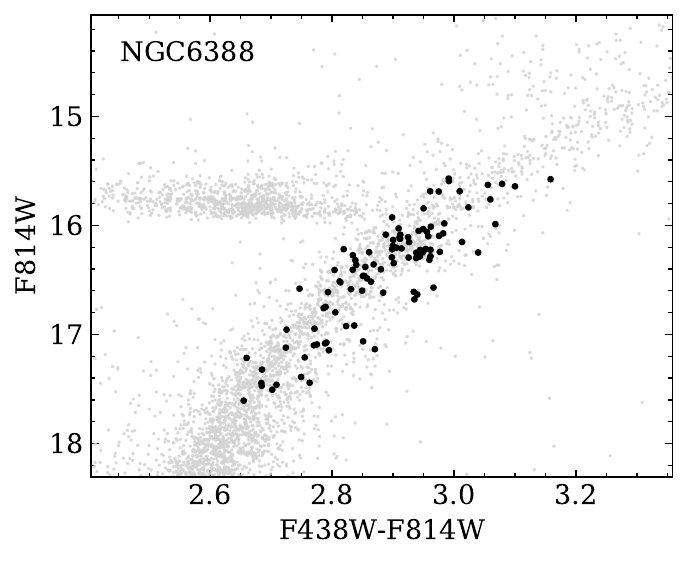}
   }  
     \caption{CMDs of NGC\,5286, NGC\,2808 and NGC\,6388. The P1 stars included in our final samples are shown with black dots. The small grey dots are the stars from the HUGS photometric catalog that have well-measured photometry (see Sect.~\ref{sec:met:cmap}).
     }
     \label{fig:cmds}
\end{figure*}

\section{Methods}\label{sec:met}

\subsection{Spectroscopic metallicities}\label{sec:met:metal}
The individual MUSE spectra of the RGB stars are fitted to derive atmospheric parameters using the G\"ottingen spectral library of \textsc{phoenix} spectra \citep{husser2013} and the fitting framework \textit{spexxy}\footnote{\url{https://github.com/thusser/spexxy}} \citep{husser2016}. Because the spectra of RGB stars are not very sensitive to changes in surface gravity (log~$g$), this parameter is obtained from an isochrone. For each GC we find an isochrone (from \citealt{marigo17}) that best reproduces the HST photometry in the F606W-F814W versus F606W CMD and we derive a value of log~$g$ and \teff\ for each star by finding the nearest point on the isochrone. This is done with two sets of photometry: the ACS Survey of globular clusters \citep{sarajedini2007,anderson08} and the $HST$ UV Globular Cluster Survey (HUGS, \citealt{piotto15,nardiello18}). The log $g$ of the star is then fixed to the average value obtained from both sets of photometry during the spectral fit. The \teff\ estimated from the isochrone is used as a starting value for the spectral fitting procedure and for selection criteria (see Sect. ~\ref{sec:met:sample}). The resulting best fit provides a value for the effective temperature (\teff), metallicity (\met), radial velocity, a model for the telluric lines and a polynomial function describing the continuum. The metallicities [M/H] of the \textsc{phoenix} model grid are solar-scaled, except for the $\alpha$-elements that are enhanced to a value that is kept fixed during the fitting procedure. The $\alpha$-enhancement used varies from cluster to cluster and is between [$\alpha$/Fe] = 0.1 and 0.4 \citep{dias2016}. 
During the spectral fit, we do not target specific spectral regions but fit the whole spectral range, except the region containing the interstellar NaD lines and the AO gap (5780-5990 \AA) in AO observations. Thus what we measure from the spectral fit is the overall metallicity \met\ and not a direct iron abundance.
A more thorough description of the spectral fitting method is included in \citet{husser2016} and \citet{nitschai23}. 

The next step is to combine, for a given star, the resulting parameters from the multiple observations to derive average values of \teff, radial velocity, and most important for this work, the metallicity (\met). The general idea is to calculate the weighted average of the individual measurements (from the individual spectra), with the weight ($w$) being the inverse of the squared uncertainty ($\epsilon$) returned by the Levenberg-Marquardt optimization routine used by \textit{spexxy}.
The resulting uncertainty on the weighted average is then: \\
\begin{math}
\sigma = \sqrt{\frac{1}{\sum w_i}}\end{math}, where \begin{math}  w = \frac{1}{\epsilon^2}
\end{math}. \\
For a spectrum to be included in the calculation of the average parameters, two quality criteria are required: the spectrum must have a signal-to-noise ratio ($S/N$) $>$ 20 and a magnitude accuracy above 0.8. To determine the magnitude accuracy parameter, \textsc{PampelMuse} calculates the differences between the magnitudes recovered from the extracted spectra and the true magnitudes available in the photometric reference catalog. A value of 1 indicates that the magnitude difference determined for a star is fully consistent with the typical differences measured for similarly bright stars, whereas 0 indicates a strong outlier (see Sect. 4.4 in \citealt{kamann2018}, for more details). 
The uncertainties on the \met\ measurements are mainly defined by two factors. First, they scale with the mean $S/N$ of the individual spectra because a higher $S/N$ spectrum results in a smaller statistical uncertainty $\epsilon$. Secondly, they depend on the number of measurements included in the average calculation.
A final factor affecting the \met\ uncertainties is the metallicity itself, meaning that the stars in low-metallicity GCs, such as NGC\,7078 and NGC\,7099, have larger uncertainties than stars at higher metallicity (like those of NGC\,104) at a given $S/N$. 
The behavior of the uncertainties (\met\ err) is illustrated in Fig.~\ref{fig:Met_SNR_err} where we show their relationship to the average $S/N$ of the stars' spectra in two GCs: NGC\,104 and NGC\,7078. For illustration purposes, we included in these figures only the P1 stars included in the analysis described below. 
NGC\,104 is relatively metal-rich ([Fe/H]=$-$0.7) and is one of the GCs with the largest number of observations, resulting in small typical uncertainties ([M/H] err $<$ 0.015 dex). NGC\,7078 is the most metal-poor GC in our sample ([Fe/H]=$-$2.4) and has fewer observations, typically below six measurements, resulting in larger uncertainties than in NGC\,104. In the case of NGC\,6362, we only have one measurement per star, so the uncertainties on [M/H] for that cluster are those returned by the Levenberg-Marquardt optimization routine of \textit{spexxy}. The median values of [M/H] err for the P1 stars of each cluster are listed in Table~\ref{table_results}.

\subsection{Photometry and chromosome maps}\label{sec:met:cmap}

The second essential ingredient for our analysis is the chromosome map of each GC. Our chromosome maps are constructed following the method described in \citet{mil17} and have been presented and used in previous work by our group \citep{Husser20,martens23,latour19}. Before making the chromosome maps, we clean the HST photometry following the procedure described in Sect. 3 of \citet{nardiello18} (see also \citealt{milone2012}). This procedure allows us to select stars with well-measured photometry according to their photometric error and the shape and quality of their point spread function during the photometric extraction. We keep only the stars that pass the selection criteria for these three parameters in all four filters necessary to create the chromosome maps. 

For our study, the chromosome maps are not necessary only to separate the stars into their respective population, but their exact pseudo-colors, especially $\Delta_{\rm F275W,F814W}$ (the x-axis of the chromosome map), are important to look for metallicity trends. One improvement we made for this work is the inclusion of the differential reddening corrections provided by \citet{Legnardi23} that is available for seven GCs in our sample (NGC\,3201, NGC\,5286, NGC\,6254, NGC\,6388, NGC\,6441, NGC\,6541, and NGC\,6656). We also paid particular care to the fiducial lines defining the RGB envelope. Particularly, we removed the brightest RGB stars, down to two magnitudes (in F814W) below the tip of the RGB. This is to avoid the issue of poorly defined fiducial lines due to the small number of stars, which causes the pseudo-colors to be more uncertain for bright RGB stars. 
From the chromosome map, we separate the stars into their respective populations (P1 and P2), which is more or less straightforward depending on the cluster. 
In the Type II GCs, we also isolate the P3 stars (those from the red-RGB), but these stars are not included in the analyses. The chromosome maps of the GCs in our study are shown in Appendix~\ref{App_B} (Figs~\ref{fig:app_cmap1}-\ref{fig:app_cmap3}) along with the P1 and P2 stars included in our samples.

\setcounter{table}{1}
\begin{table*}[t]
\caption{Parameters of the $\Delta_{\rm F275W,F814W}$ pseudo-color$-$metallicity relationship and metallicity spread derived among the P1 stars. }
\label{table_results}      
\centering                    
\begin{tabular}{l c c c c c c c c c}        
\toprule\toprule
\noalign{\vskip4bp}
Cluster &  $N$star & [M/H]err & $\Delta C$ & $R_{\rm P}$ & $p$-value & $a$ & $b$ & $\Delta$\met & \\
 & (1) & (2) & (3) & (4) & (5) & (6) & (7) & (8) \\
 \midrule
NGC\,104 & 226 & 0.008 & 0.29 & 0.58 & 7.96e-22 &  0.21 $\pm$ 0.03 & -0.787 $\pm$ 0.003 & 0.061 $\pm$ 0.008 \\
NGC\,1851 & 127 & 0.011 & 0.21 & 0.40 & 4.29e-06 &  0.25 $\pm$ 0.10 & -1.169 $\pm$ 0.011 & 0.053 $\pm$ 0.016 \\
NGC\,2808 & 152 & 0.018 & 0.33 & 0.57 & 3.35e-14 &  0.41 $\pm$ 0.08 & -1.072 $\pm$ 0.008 & 0.137 $\pm$ 0.021 \\
NGC\,3201 & 30 & 0.007 & 0.25 & 0.69 & 2.29e-05 &  0.38 $\pm$ 0.15 & -1.449 $\pm$ 0.018 & 0.095 $\pm$ 0.029 \\
NGC\,362 & 113 & 0.017 & 0.12 & 0.21 & 2.54e-02 &  0.32 $\pm$ 0.20 & -1.159 $\pm$ 0.009 & 0.039 $\pm$ 0.020 \\
NGC\,5286 & 115 & 0.023 & 0.29 & 0.67 & 3.27e-16 &  0.51 $\pm$ 0.10 & -1.585 $\pm$ 0.016 & 0.148 $\pm$ 0.023 \\
NGC\,5904 & 113 & 0.015 & 0.20 & 0.63 & 7.36e-14 &  0.41 $\pm$ 0.09 & -1.296 $\pm$ 0.006 & 0.083 $\pm$ 0.015 \\
NGC\,6093 & 234 & 0.025 & 0.20 & 0.24 & 2.02e-04 &  0.15 $\pm$ 0.09 & -1.647 $\pm$ 0.008 & 0.030 $\pm$ 0.014 \\
NGC\,6218 & 64 & 0.015 & 0.13 & 0.22 & 7.55e-02 &  0.17 $\pm$ 0.18 & -1.319 $\pm$ 0.011 & 0.022 $\pm$ 0.018 \\
NGC\,6254 & 83 & 0.016 & 0.23 & 0.68 & 1.51e-12 &  0.48 $\pm$ 0.13 & -1.497 $\pm$ 0.009 & 0.110 $\pm$ 0.023 \\
NGC\,6362 & 33 & 0.025 & 0.21 & 0.38 & 2.76e-02 &  0.24 $\pm$ 0.17 & -1.090 $\pm$ 0.013 & 0.050 $\pm$ 0.028 \\
NGC\,6388 & 97 & 0.018 & 0.80 & 0.81 & 1.45e-23 &  0.30 $\pm$ 0.04 & -0.471 $\pm$ 0.012 & 0.240 $\pm$ 0.025 \\
NGC\,6441 & 148 & 0.018 & 0.48 & 0.35 & 1.34e-05 &  0.14 $\pm$ 0.07 & -0.406 $\pm$ 0.010 & 0.067 $\pm$ 0.026 \\
NGC\,6541 & 255 & 0.017 & 0.13 & 0.16 & 1.12e-02 &  0.11 $\pm$ 0.08 & -1.727 $\pm$ 0.006 & 0.014 $\pm$ 0.009 \\
NGC\,6624 & 68 & 0.017 & 0.36 & 0.48 & 3.33e-05 &  0.15 $\pm$ 0.09 & -0.761 $\pm$ 0.013 & 0.054 $\pm$ 0.025 \\
NGC\,6656 & 72 & 0.014 & 0.23 & 0.55 & 5.14e-07 &  0.56 $\pm$ 0.17 & -1.715 $\pm$ 0.035 & 0.132 $\pm$ 0.032 \\
NGC\,6681 & 38 & 0.024 & 0.17 & 0.02 & 8.99e-01 &  -0.05 $\pm$ 0.22 & -1.546 $\pm$ 0.014 & -0.008 $\pm$ 0.028 \\
NGC\,6752 & 92 & 0.012 & 0.16 & 0.72 & 5.50e-16 &  0.71 $\pm$ 0.11 & -1.499 $\pm$ 0.006 & 0.117 $\pm$ 0.014 \\
NGC\,7078 & 220 & 0.033 & 0.17 & 0.47 & 2.86e-13 &  0.71 $\pm$ 0.15 & -2.195 $\pm$ 0.010 & 0.122 $\pm$ 0.021 \\
NGC\,7089 & 171 & 0.019 & 0.24 & 0.70 & 1.23e-26 &  0.66 $\pm$ 0.09 & -1.484 $\pm$ 0.008 & 0.156 $\pm$ 0.018 \\
NGC\,7099 & 67 & 0.021 & 0.10 & 0.05 & 6.63e-01 &  0.22 $\pm$ 0.45 & -2.172 $\pm$ 0.020 & 0.021 $\pm$ 0.034 \\
\bottomrule
\end{tabular}
\vspace{0.9em}
\parbox{\textwidth}{\small Notes. (1) Number of stars included in the P1 sample. (2) Median value of the errors on [M/H]. (3) \cmapx\ spread of the P1 stars. (4)~Pearson correlation coefficient between \cmapx\ and [M/H] and its (5) $p$-value. (6) $a$ and (7) $b$ coefficients from the linear relationship derived as in \cmapx = $a$[M/H] + $b$ and their 95\% confidence interval as uncertainties. (8) Metallicity spread. }
\end{table*}

\subsection{Sample selection}\label{sec:met:sample}
Our selection of stars from the chromosome maps is then matched with our sample of MUSE stars with \met\ measurements. We apply several additional criteria to define our final sample of stars with the most reliable spectroscopic metallicity measurements and $\Delta_{\rm F275W,F814W}$ position. We enumerate below the selection criteria.

\begin{itemize}

    \item We keep stars having at least three [M/H] measurements. The two exceptions are NGC\,6681 and NGC\,6362. For NGC\,6681 we only have two observations of a single central pointing; in that case, we keep the stars that have two measurements. For NGC\,6362 there is only one spectrum per star so we keep all stars.
    
    \item We remove stars having a probability of radial velocity variability above 0.8, as defined in Sect. 5 of \citet{Giesers2019}. This means that we remove from our sample the stars showing radial velocity variations, most likely due to binarity. Binaries were found to populate, among others, the blue extension of the $\Delta_{\rm F275W,F814W}$ pseudo-color in the chromosome map \citep{Kamann2020,Martins2020}. While this is efficient at removing binaries in clusters with many observations, such as NGC\,3201 and NGC\,104, there are likely still binaries among the stars with a small number of observations. We note that this criteria is not applied to NGC\,6362 because we have only one measurement per star.

    \item We keep only stars for which the difference between the \teff\ estimated from the isochrone and the spectroscopic \teff\ is within 3$\sigma$ of the average \teff\ difference between isochrone and spectroscopy. Effectively, this removes stars whose spectra returned a \teff\ that is discrepant from their photometric properties (i.e., their position along the RGB). This can happen, for example, if the star is severely blended with a close neighbor causing contamination in the extracted spectra. This is done on a cluster-by-cluster basis.

    \item Similarly, we keep only stars for which the difference between the log $g$ returned from both sets of photometry (ACS and HUGS) is within 3$\sigma$ of the average difference. Effectively, this removes stars that have discrepant photometry between ACS and HUGS. If their position on both CMDs is different, it results in different log $g$ estimated from the isochrones. This is done on a cluster-by-cluster basis.

    \item We remove the brightest RGB stars, one to two magnitudes below the brightest star, depending on the cluster. The cut is made depending on how populated is the RGB because with fewer stars, the fiducial lines, thus the $\Delta_{\rm F275W,F814W}$ position is more uncertain. This cut also prevents contamination by AGB stars because they are found along the brightest part of the RGB. Finally, by limiting the range of magnitudes spanned by our stars, we also limit the extent of the metallicity-magnitude trend that affects the data (this is discussed in Appendix~\ref{App_A}).

    \item We cross-match the stars with the HST proper-motion catalogs of \citet{Libralato2022} to remove any star that might not be a cluster member from its proper motion.    
    
    \item We remove stars with a membership probability $p_{\rm member}<$ 0.5, as defined in \citet{kamann2016}. Our membership probability is based on a combination of radial velocity and metallicity. While this is mostly redundant with the proper motion selection, it is still relevant for bright stars because some of them do not have proper motion measurements.
    
\end{itemize}

After applying this selection, we have between 30 and 250 P1 stars per cluster. The majority of them are fainter than the horizontal branch. In Fig.~\ref{fig:cmds} we show the position of the P1 stars in the $F438W$ versus $F814W$ CMD for three example clusters: NGC\,5286, NGC\,2808 and NGC\,6388. 
For each star in our final samples, from P1 and P2, we make available their atmospheric parameters (\met, \teff, and, log $g$) and their pseudo-colors from our chromosome maps in Table 1 that is only available online at CDS (see Sect.\ref{data_availability}).

\subsection{Metallicity variation and dispersion}\label{sec:met:var}

To estimate the metallicity spread among the P1 stars of each cluster, we perform a linear regression in the $\Delta_{\rm F275W,F814W}$ $-$ \met\ plane. Because the individual measurements have various precision, depending on their errors, we take that into account by performing a weighted least-square (WLS) regression, using the inverse of the squared error as weight. 
From the relationship obtained, we can estimate the metallicity variation for a given range of $\Delta_{\rm F275W,F814W}$ pseudo-color.
To compute the pseudo-color extension ($\Delta C$) of the P1 stars in each GC, we consider the range between the 4$^{th}$ and 96$^{th}$ percentile of the P1 stars in our final sample. The \dc\ obtained from our spectroscopic sample is representative of that of the whole photometric sample in the chromosome maps because, as opposed to the previous studies using high-resolution spectroscopy (e.g. \citealt{Lardo18,Marino23_47tuc}), we do not have any pre-selection of the targets based on their position in the chromosome map. 
From the linear relation obtained with the WLS regression, we compute the metallicity predicted at the 4$^{th}$ and 96$^{th}$ pseudo-color percentiles and calculate the metallicity variation as $\Delta$\met~=~\met$_{96 \rm th}$~$-$~\met$_{4 \rm th}$. We can consider here two different values as uncertainties, either from the standard errors returned on the parameters of the linear equation or from the 95\% confidence interval of the regression. To be more conservative, in the following we compute and list uncertainties considering the 95\% confidence interval.

We also compute the metallicity dispersion among our samples of stars. 
This is done using a Markov chain Monte Carlo ensemble sampler developed by \citet{emcee2013} to model our metallicity measurements with a Gaussian function. We then retrieved the most likely average metallicity and standard deviation ($\sigma$), with their uncertainties being the 16$^{th}$ and 84$^{th}$ percentile of the posterior probability distributions (equivalent to 1$\sigma$). This method takes into account the uncertainties on \met, thus returning a standard deviation that should be representative, if the uncertainties are not significantly underestimated, of the intrinsic dispersion. We do not claim to measure an accurate intrinsic metallicity dispersion from our data, but we believe our values provide robust upper limits.

\begin{figure*}
\resizebox{\hsize}{!}{
   \includegraphics{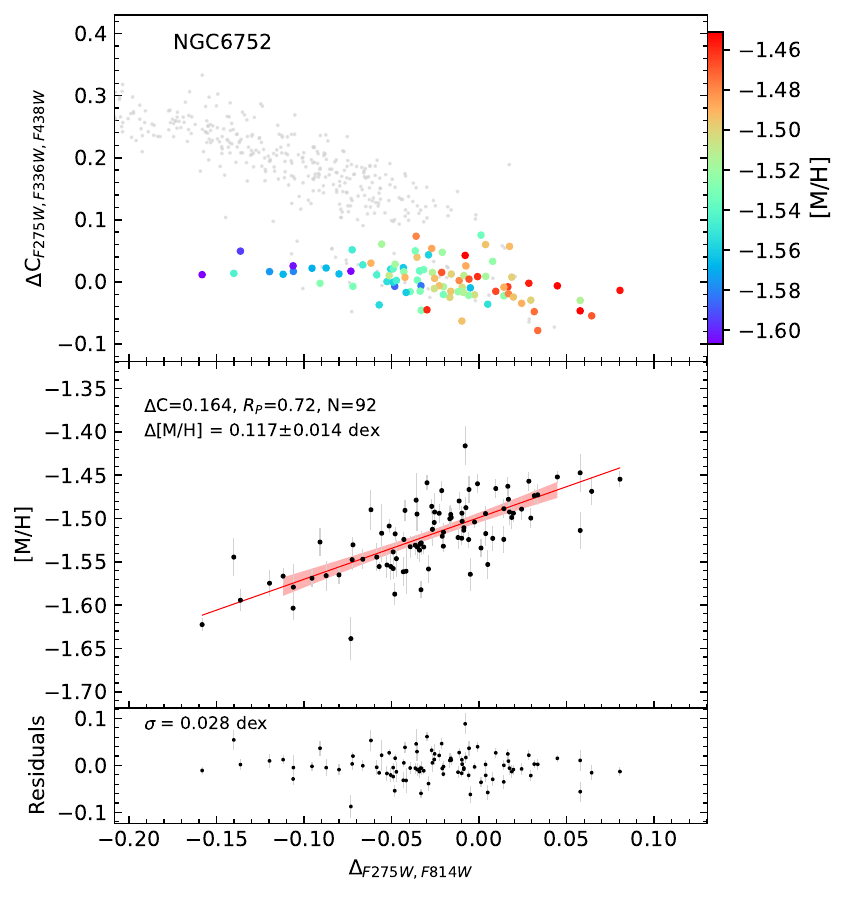}
   \includegraphics{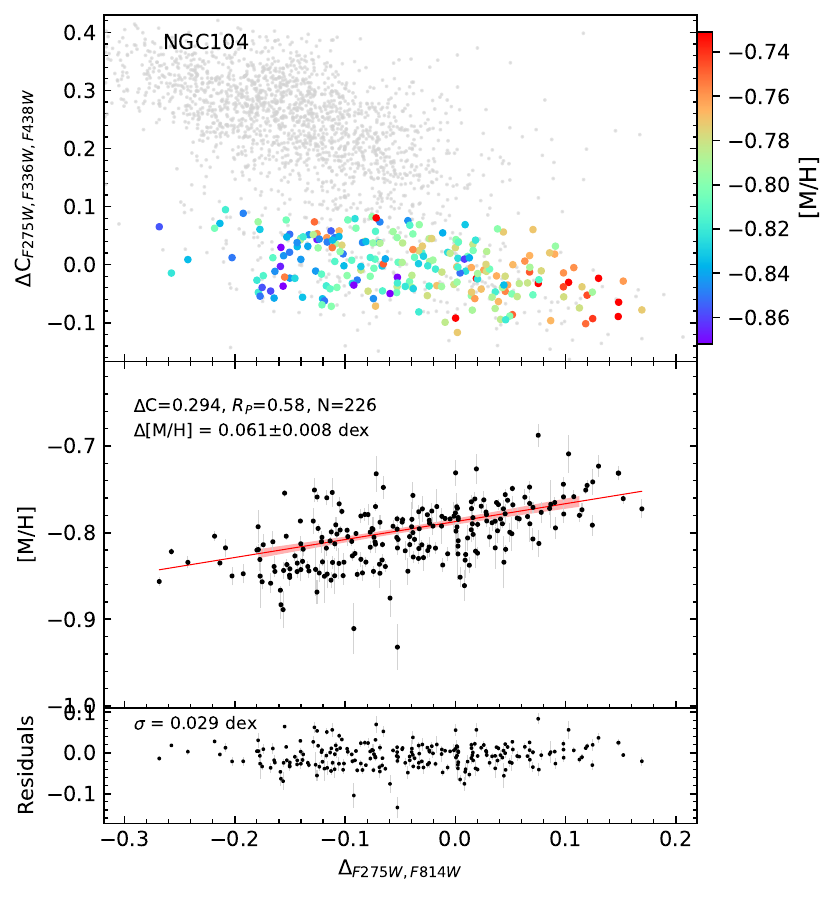}}\vspace{1pt}
\resizebox{\hsize}{!}{   
   \includegraphics{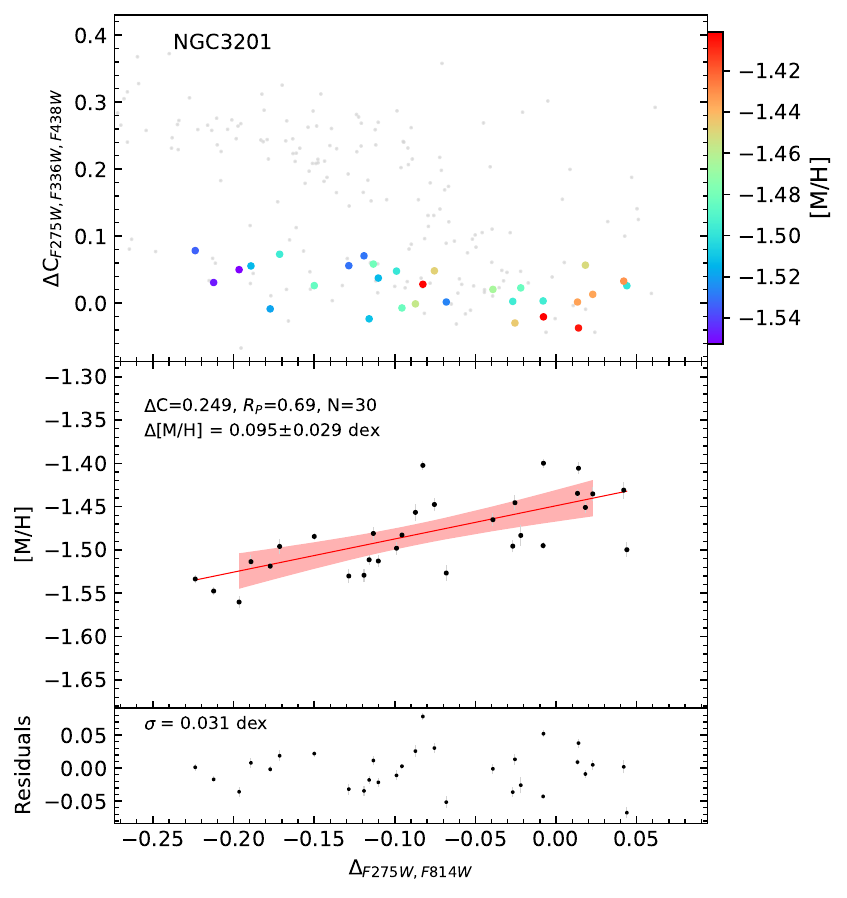}
   \includegraphics{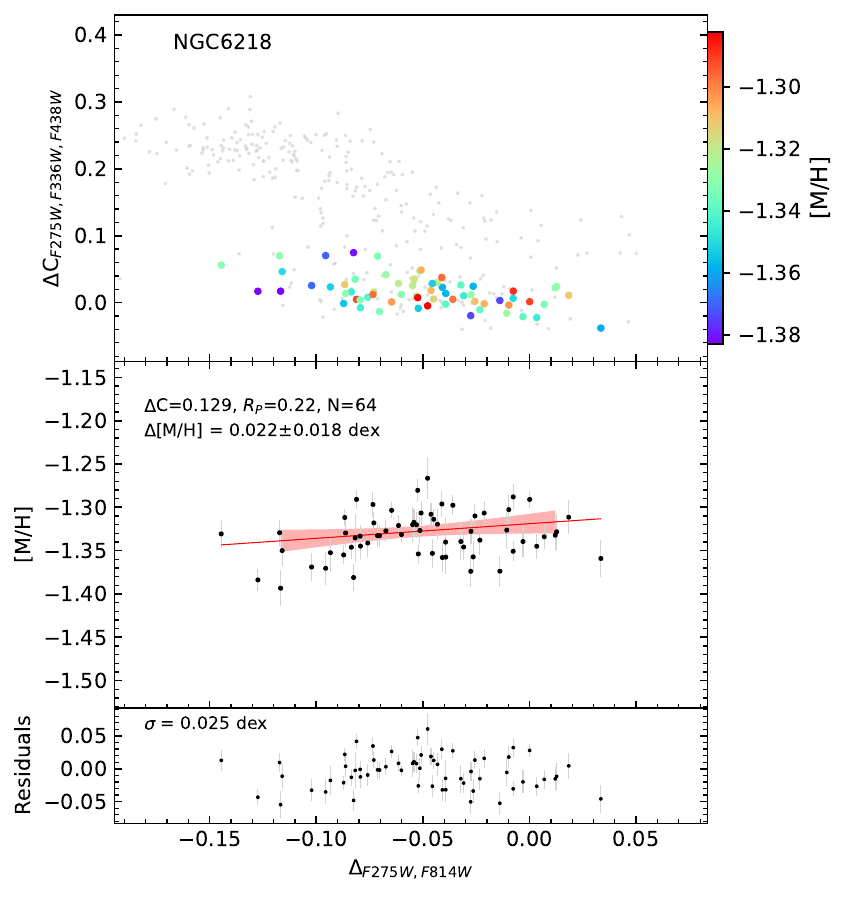}   
   }
     \caption{Metallicity \met\ versus $\Delta_{\rm F275W,F814W}$ pseudo-color relationship for four GCs. On the top panels, we show the chromosome maps of the clusters with the P1 stars within our sample color-coded by their metallicity. The middle panels show the metallicity of each star, with the relationship derived from the WLS regression (red line) and the 95\% confidence interval (red shaded area) over the color range \dc. We also indicate the number of stars ($N$), the Pearson correlation coefficient ($R_{\rm P}$), and the resulting metallicity variation $\Delta$\met. The bottom panels show the residuals as in observed$-$predicted and we indicate the standard deviation ($\sigma$) of the residuals. 
     }
     \label{fig:plots}
\end{figure*}

\begin{figure*}
\resizebox{\hsize}{!}{
   \includegraphics{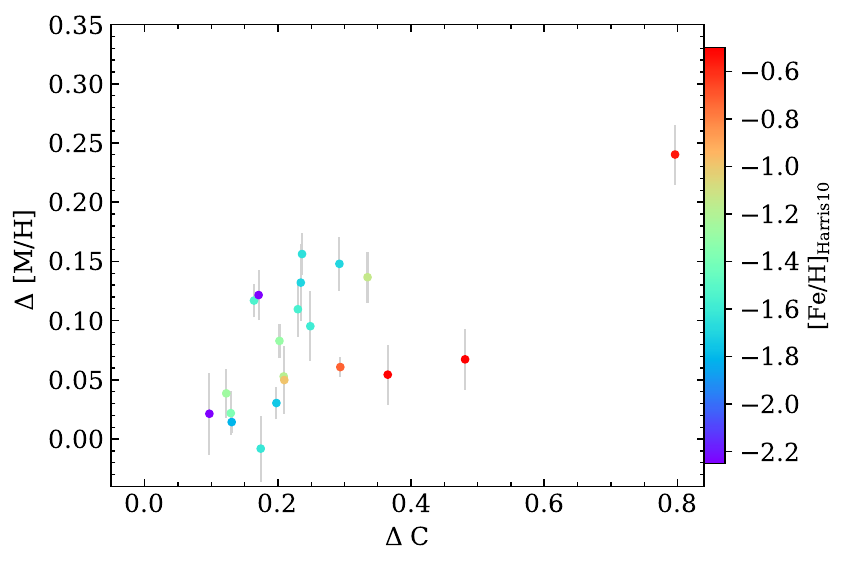}
   \includegraphics{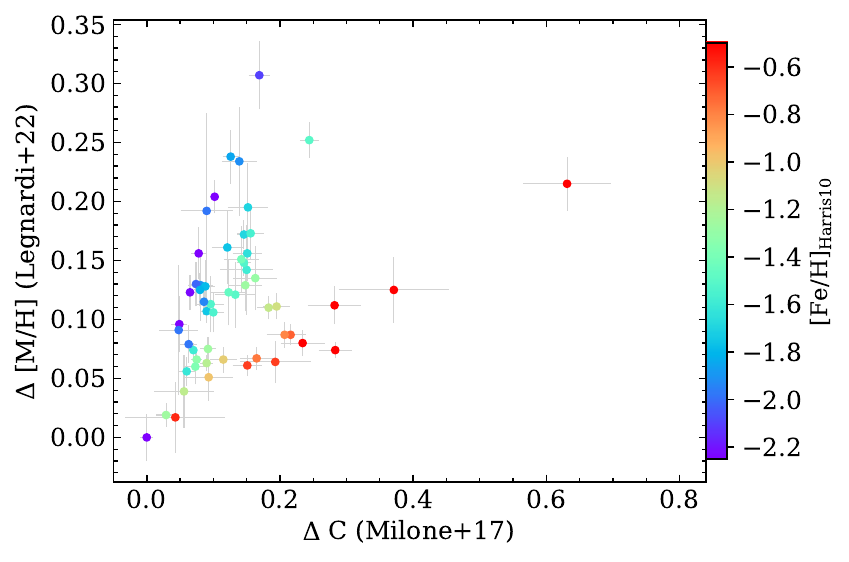}}  
     \caption{Metallicity variations within the P1 stars versus their color extension for the GC included in our study (left panel) and in the study of \citet{Legnardi22} (right panel). Each point represents a GC and is color-coded by the average metallicity of the cluster (\citeauthor[2010 edition]{harris1996}). }
     \label{fig:dC_dM}
\end{figure*}

\section{Results}\label{sec:res}

\subsection{Metallicity spread among the P1 stars}\label{sec:res:met_spread}

Our results are summarized in Table \ref{table_results}. For each GC we list the number of P1 stars included (Nstar), the median uncertainty on the stars' metallicity ([M/H]err) as an indicator of our measurements precision, the range in pseudo-color spanned by the P1 stars (\dc, as described in Sect. 3.4), the Pearson correlation coefficient $R_{\rm P}$ and its associated $p$-value. We use the Pearson correlation coefficient because we expect a linear relationship. We also include the $a$ and $b$ coefficients of the relationship derived, as in $y = ax+b$, and their 95\% confidence interval as uncertainties. Finally, we list the metallicity variations $\Delta$\met\ with their uncertainties.
The first thing to examine is whether we find a significant correlation between metallicity and color. This is the case for 17 out of the 21 GCs. We measure a slope $a$ that is consistent with zero, only for three clusters, namely NGC\,7099, NGC\,6681, and NGC\,6218. These three clusters have a $p$-value larger than 7\%. For all other clusters, the slope and $R_{\rm P}$ are positive, which demonstrates the sensitivity of the $\Delta_{\rm F275W,F814W}$ pseudo-color to changes in metallicity. We illustrate our results in Fig.~\ref{fig:plots} for four clusters: NGC\,6752, NGC\,104, NGC\,3201, and NGC\,6218. The results for the other clusters are shown in the Appendix (Fig.~\ref{fig:app1} to~\ref{fig:app5}).  

The metallicity variations $\Delta$\met\ that we obtain are between $\sim$ 0.03 and 0.2 dex. In the hypothesis that the color extension of P1 is caused by metallicity variations, we expect a correlation between \dc\ and $\Delta$\met. This is illustrated in Fig.~\ref{fig:dC_dM}. However, the relationship is not strictly linear but also shows a dependency on the average metallicity of the GC in the sense that for a given \dc, we find the higher metallicity GCs 
to have smaller metallicity variations within their P1 stars. This behavior is well illustrated from the results of \citet{Legnardi22} shown on the right panel of Fig.~\ref{fig:dC_dM}.
This clear dependence between the metallicity spread ($\Delta$\met) and the GC average metallicity is a natural consequence of the photometric method used by \citet{Legnardi22} that relies on the color-metallicity relation of isochrones; a given metallicity increase, say by 0.05 dex, has a smaller impact on the F275W magnitude, where most iron and heavy-metal lines are found, at low metallicity than at larger ones. 
This dependency is also visible in our data for the higher metallicity clusters.
We note here that even though the \citet{Legnardi22} results are referred to as iron variations ($\Delta$[Fe/H]), their isochrones were generated with different metallicities and not only with different iron abundances (Legnardi, private comm. 2024). We thus take the liberty to re-label their results as metallicity variations ($\Delta$\met).
 
It is worth mentioning here that NGC\,6388, with its particularly large \dc\ of 0.8, is not part of the \citet{Legnardi22} sample because there is no estimate of a color spread for the P1 stars of that cluster. In any case, the RGB of NGC\,6388 is especially wide in the 
$F275W$-$F814W$ color (see \citealt{mil17}, but also \citealt{carretta2022_ngc6388}) and the pseudo-color $-$ metallicity relation is conspicuous from Fig.~\ref{fig:app2}, this results in our largest $\Delta$\met\ of 0.24 dex.

\subsection{Metallicity dispersion}\label{sec:res:sigmas}

\begin{table*}[h]
\small
\caption{Mean metallicity and dispersion measured in the P1 and P2 stars with their 1$\sigma$ uncertainties.}
\label{table_sigmas}      
\centering                    
\begin{tabular}{l c c c c c c c c c c}        
\toprule\toprule
\noalign{\vskip4bp}
Cluster & $N$stars & Mean [M/H] & $\sigma$ [M/H] & & $N$stars & Mean [M/H] & $\sigma$ [M/H] & & Mean [M/H] & $\sigma$ [M/H] \\ 
& \multicolumn{3}{c}{P1} & & \multicolumn{3}{c}{P2} & & \multicolumn{2}{c}{P1+P2} \\
 \cline{2-4} \cline{6-8} \cline{10-11}
 \noalign{\vskip3bp}
NGC\,104 & 226 & -0.802 $\pm$ 0.002 & 0.033 $\pm$ 0.002 & & 916 & -0.802 $\pm$ 0.001 & 0.032 $\pm$ 0.001 & & -0.802 $\pm$ 0.001 & 0.033 $\pm$ 0.001 \\ 
NGC\,1851 & 127 & -1.207 $\pm$ 0.003 & 0.036 $\pm$ 0.003 & & 239 & -1.210 $\pm$ 0.002 & 0.031 $\pm$ 0.002 & & -1.209 $\pm$ 0.002 & 0.033 $\pm$ 0.001 \\ 
NGC\,2808 & 152 & -1.108 $\pm$ 0.005 & 0.056 $\pm$ 0.004 & & 483 & -1.096 $\pm$ 0.002 & 0.052 $\pm$ 0.002 & & -1.105 $\pm$ 0.003 & 0.055 $\pm$ 0.003 \\ 
NGC\,3201 & 30 & -1.482 $\pm$ 0.008 & 0.045 $\pm$ 0.006 & & 47 & -1.474 $\pm$ 0.004 & 0.029 $\pm$ 0.003 & & -1.477 $\pm$ 0.004 & 0.036 $\pm$ 0.003 \\ 
NGC\,362 & 113 & -1.178 $\pm$ 0.004 & 0.038 $\pm$ 0.003 & & 281 & -1.195 $\pm$ 0.002 & 0.037 $\pm$ 0.002 & & -1.191 $\pm$ 0.002 & 0.038 $\pm$ 0.002 \\ 
NGC\,5286 & 115 & -1.663 $\pm$ 0.006 & 0.057 $\pm$ 0.005 & & 174 & -1.654 $\pm$ 0.004 & 0.046 $\pm$ 0.003 & & -1.658 $\pm$ 0.003 & 0.051 $\pm$ 0.003 \\ 
NGC\,5904 & 113 & -1.307 $\pm$ 0.004 & 0.037 $\pm$ 0.003 & & 355 & -1.315 $\pm$ 0.001 & 0.022 $\pm$ 0.001 & & -1.313 $\pm$ 0.001 & 0.027 $\pm$ 0.001 \\ 
NGC\,6093 & 234 & -1.661 $\pm$ 0.003 & 0.034 $\pm$ 0.003 & & 324 & -1.662 $\pm$ 0.002 & 0.027 $\pm$ 0.002 & & -1.662 $\pm$ 0.002 & 0.030 $\pm$ 0.002 \\ 
NGC\,6218 & 64 & -1.330 $\pm$ 0.003 & 0.021 $\pm$ 0.003 & & 97 & -1.345 $\pm$ 0.003 & 0.022 $\pm$ 0.002 & & -1.339 $\pm$ 0.002 & 0.022 $\pm$ 0.002 \\ 
NGC\,6254 & 83 & -1.519 $\pm$ 0.005 & 0.047 $\pm$ 0.004 & & 171 & -1.524 $\pm$ 0.002 & 0.026 $\pm$ 0.002 & & -1.523 $\pm$ 0.002 & 0.034 $\pm$ 0.002 \\ 
NGC\,6362 & 33 & -1.101 $\pm$ 0.006 & 0.024 $\pm$ 0.007 & & 18 & -1.119 $\pm$ 0.007 & 0.012 $\pm$ 0.009 & & -1.108 $\pm$ 0.005 & 0.020 $\pm$ 0.006 \\ 
NGC\,6388 & 97 & -0.541 $\pm$ 0.008 & 0.078 $\pm$ 0.006 & & 312 & -0.498 $\pm$ 0.004 & 0.064 $\pm$ 0.003 & & -0.507 $\pm$ 0.004 & 0.070 $\pm$ 0.003 \\ 
NGC\,6441 & 148 & -0.433 $\pm$ 0.005 & 0.055 $\pm$ 0.004 & & 374 & -0.378 $\pm$ 0.004 & 0.083 $\pm$ 0.003 & & -0.394 $\pm$ 0.004 & 0.080 $\pm$ 0.003 \\ 
NGC\,6541 & 255 & -1.736 $\pm$ 0.002 & 0.026 $\pm$ 0.002 & & 242 & -1.743 $\pm$ 0.002 & 0.027 $\pm$ 0.002 & & -1.740 $\pm$ 0.001 & 0.028 $\pm$ 0.001 \\ 
NGC\,6624 & 68 & -0.786 $\pm$ 0.005 & 0.038 $\pm$ 0.004 & & 175 & -0.804 $\pm$ 0.003 & 0.029 $\pm$ 0.002 & & -0.799 $\pm$ 0.002 & 0.033 $\pm$ 0.002 \\ 
NGC\,6656 & 72 & -1.831 $\pm$ 0.007 & 0.059 $\pm$ 0.005 & & 107 & -1.806 $\pm$ 0.006 & 0.065 $\pm$ 0.005 & & -1.816 $\pm$ 0.005 & 0.064 $\pm$ 0.004 \\ 
NGC\,6681 & 38 & -1.547 $\pm$ 0.005 & 0.024 $\pm$ 0.005 & & 180 & -1.558 $\pm$ 0.003 & 0.032 $\pm$ 0.003 & & -1.556 $\pm$ 0.003 & 0.031 $\pm$ 0.002 \\ 
NGC\,6752 & 92 & -1.518 $\pm$ 0.004 & 0.038 $\pm$ 0.003 & & 220 & -1.514 $\pm$ 0.002 & 0.027 $\pm$ 0.002 & & -1.517 $\pm$ 0.002 & 0.032 $\pm$ 0.002 \\ 
NGC\,7078 & 220 & -2.242 $\pm$ 0.005 & 0.067 $\pm$ 0.004 & & 329 & -2.200 $\pm$ 0.003 & 0.053 $\pm$ 0.003 & & -2.237 $\pm$ 0.004 & 0.064 $\pm$ 0.003 \\ 
NGC\,7089 & 171 & -1.515 $\pm$ 0.005 & 0.058 $\pm$ 0.004 & & 670 & -1.526 $\pm$ 0.002 & 0.038 $\pm$ 0.001 & & -1.524 $\pm$ 0.002 & 0.044 $\pm$ 0.001 \\ 
NGC\,7099 & 67 & -2.184 $\pm$ 0.007 & 0.051 $\pm$ 0.005 & & 150 & -2.167 $\pm$ 0.003 & 0.028 $\pm$ 0.003 & & -2.172 $\pm$ 0.003 & 0.038 $\pm$ 0.002 \\ 

\bottomrule
\end{tabular}
\end{table*}

\begin{figure}
\resizebox{\hsize}{!}{
   \includegraphics{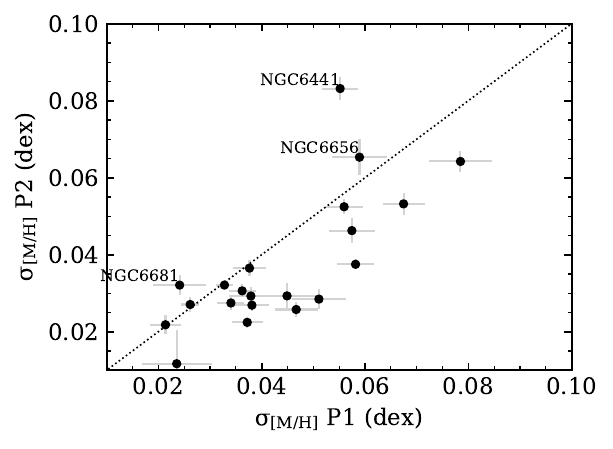}}
     \caption{Metallicity dispersion of the P2 stars versus that of the P1 stars. The error bars are 1$\sigma$ uncertainties. The identity relation is drawn with a dashed line. The three GCs for which the P2 stars have a larger dispersion than the P1 stars are indicated with their name.
     }
     \label{fig:sigmaP1P2}
\end{figure}

Table~\ref{table_sigmas} presents the average and dispersion values of the metallicities \met\ obtained from our samples of P1 and P2 stars. The respective uncertainties correspond to 1$\sigma$. The number of stars included in each sample is also indicated. The P2 stars were first selected from their position in the chromosome maps (see Figs.~\ref{fig:app_cmap1}-\ref{fig:app_cmap3}), and then filtered according to the criteria listed in Sect.~\ref{sec:met:sample}. 
The metallicity dispersion of the P2 stars was computed in the same way as that of the P1 stars.

While we know that the chemical composition, at least in terms of light elements, is homogeneous in the P1 stars, this is not the case for the P2 stars. This could potentially affect our spectral fitting procedure because the \textsc{phoenix} models do not account for variations of individual atomic species like Na, O, Mg, and Al. These four elements have spectral lines, in the MUSE spectra, and their strength varies between populations (see, e.g., \citealt{latour19} for line strength variations between the populations of NGC\,2808). Admittedly, the wavelength ranges affected by these lines are small compared to the whole spectral coverage, but they could affect the metallicity of the best-fitting model. For example, if strong lines (essentially the magnesium triplet) are significantly stronger or weaker than the model predictions, it could influence the $\chi^2$ minimum towards a higher or lower \met\ value\footnote{We note here that the strong sodium lines from the NaD doublet are always masked during the spectral fit because they are blended with the lines from the interstellar medium. Thus they do not influence the resulting metallicity.}. This could cause some small metallicity spread among the P2 stars.

We show the measured dispersion of the P2 stars versus that of the P1 stars in Fig.~\ref{fig:sigmaP1P2}. In all but three clusters, the dispersion of the P2 stars is equal to or smaller than that of the P1 stars. If systematic effects were present in our MUSE sample, we would expect them to increase the metallicity dispersion of the P2 stars, as discussed above. The fact that we measure, in most clusters, a smaller metallicity dispersion among the P2 stars, is in line with the results of \citet{Legnardi22} who found that the P1 main-sequence stars have a wider $F275W$-$F814W$ color spread than the P2 main-sequence stars in two particular GCs. The authors concluded that the metallicity spread among the P2 stars is lower than the metallicity spread of the P1 stars.
Our results show that, at least for RGB stars, this conclusion can be extended to the massive GCs of the Milky Way. We note that this important result remains unchanged by the use of metallicities corrected for the magnitude (or \teff) trend discussed in App.~\ref{App_A} (see Fig.~\ref{fig:App_sigmaP1P2corr}).
The three GCs for which we find a larger dispersion in P2 are NGC\,6681, NGC\,6656, and NGC\,6441. In the former two, the differences between $\sigma_{\rm P2}$ and $\sigma_{\rm P1}$ is less than 2$\sigma$. In NGC\,6656, there is one P2 star that is an outlier in terms of \met\ (see Fig.~\ref{fig:app_tefftrend}), removing this object lowers $\sigma_{\rm P2}$ very close to the value of $\sigma_{\rm P1}$.
In NGC\,6441, the difference is significant, but the separation of the populations is highly uncertain. The chromosome map of that cluster is peculiar and the RGB stars are essentially distributed continuously along a diagonal sequence so that it is not possible to see a distinction between P1 and P2 stars (see Fig.~\ref{fig:app_cmap3}), this is why \citet{mil17} did not estimate a $N_{\rm P1}/N_{\rm tot}$ ratio for NGC\,6441.
Our separation simply aimed at having a typical ratio of $N_{\rm P1}/N_{\rm tot} \approx 0.3$ \citep{mil17}.

\section{Discussion}\label{sec:diss}
\subsection{Comparison with literature results}

\begin{figure*}
\resizebox{\hsize}{!}{
   \includegraphics{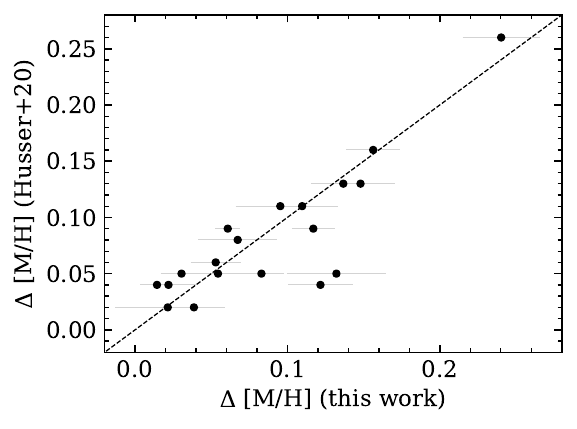}
   \includegraphics{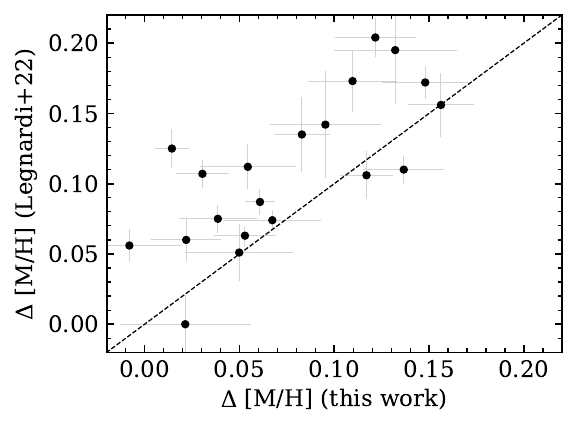}}
     \caption{Comparison between the $\Delta$\met\ found in this study and the $\Delta$[Fe/H] obtained by \citet{Husser20} (left) and \citet{Legnardi22} (right).
     The identity relation is drawn with the dashed line.}
     \label{fig:dM_vs_dFe}
\end{figure*}

As mentioned in the introduction, iron spreads among the P1 stars of GCs have been estimated from high-resolution spectroscopic measurements in only three GCs. We recall here that studies done with high-resolution spectroscopy are directly measuring iron abundances.
Nonetheless, from the abundances of numerous atomic species in the P1 stars of NGC\,104 and NGC\,3201 it appeared that not only iron is varying but also the other elements studied \citep{Marino19_3201,Marino23_47tuc}. A similar conclusion was made by \citet{Lardo2023} from their five P1 stars in NGC\,2808. It is thus reasonable to assume that iron and metallicity variations go hand in hand.
For NGC\,2808, the linear relationship shown in Fig.~7 of \citet{Lardo2023}, indicates an iron variation of about 0.15~dex among five P1 stars.
For this cluster, we computed a $\Delta$\met\ of 0.14$\pm$0.02 dex for a similar \dc\ as that of \citet{Lardo2023}.
In NGC\,3201, \citet{Marino19_3201} reported a difference of 0.1 dex among a dozen of P1 stars spanning the same \dc\ as in our sample. This is in perfect agreement with our $\Delta$\met\ of 0.095$\pm$0.03 dex. 
Finally, for NGC\,104, we used our analysis method on the literature data (color and [Fe/H] for 21 P1 stars, \citealt{Marino23_47tuc}) and obtained a $\Delta$[Fe/H] = 0.14$\pm$0.07 dex for a \dc\ of 0.22. This is larger than our value of $\Delta$\met$=0.061 \pm 0.01$ dex, even though we have a larger color range (\dc$=0.3$). 
We note that because of its relatively high metallicity ([M/H] = $-$0.8), NGC\,104 has a smaller $\Delta$\met\ than other more metal-poor clusters with a similar \dc\ value in our sample (see also Fig.~\ref{fig:dC_dM}). 

The MUSE spectra have also been used by \citet{Husser20} to estimate the metallicity spread among the P1 stars on the RGB. In that work, the metallicities were estimated from an empirical calibration of the calcium triplet (CaT) $-$ metallicity relationship. This means that the equivalent widths of the three \ion{Ca}{ii} lines ($\lambda\lambda$8498, 8542, 8662) and the stars' magnitude were used to derive the metallicity. Although the \citet{Husser20} study is also based on the MUSE spectra, the method used is independent of ours because the metallicity derived from the CaT does not rely on any model atmospheres or atmospheric parameters such as \teff\ and log $g$. The chromosome maps used in \citet{Husser20} are slightly different than those presented here (see Sect.~\ref{sec:met:cmap}) and their analysis included stars almost up to the tip of the RGB. Besides a membership criteria, there was no further "cleaning" of the stellar sample. They calculated a metallicity spread in a similar way as done here, by computing the slope of the P1 stars in the [Fe/H]$-$\cmapx\ plane (see their Fig. 21) but they used a smaller $\Delta$C than in this work to compute their metallicity spread. In the left panel of Fig~\ref{fig:dM_vs_dFe} we show a comparison between our results and those of \citet{Husser20}. Here again, we re-label their [Fe/H] for [M/H] because the calibration of the CaT equivalent widths is made using the average metallicities of GCs. 
The agreement is very good, only for two clusters, namely NGC\,6656 and NGC\,7078, do we derive a significantly larger metallicity spread.

We also compared our results with those of \citet{Legnardi22} for the 20 GCs in common (NGC\,6388 is excluded) in the right panel of Fig.~\ref{fig:dM_vs_dFe}. As explained above,
the authors used a different method to estimate metallicity variations, relying on photometry and isochrones instead of direct metallicity measurements and this results in a systematic shift between both sets of results. In general, the metallicity spreads obtained from photometry are larger than our spectroscopic measurements, but they correlate well. At this point we do not know why the photometric method results in larger metallicity spread than our spectroscopic values.

\begin{figure*}
\resizebox{\hsize}{!}{
   \includegraphics{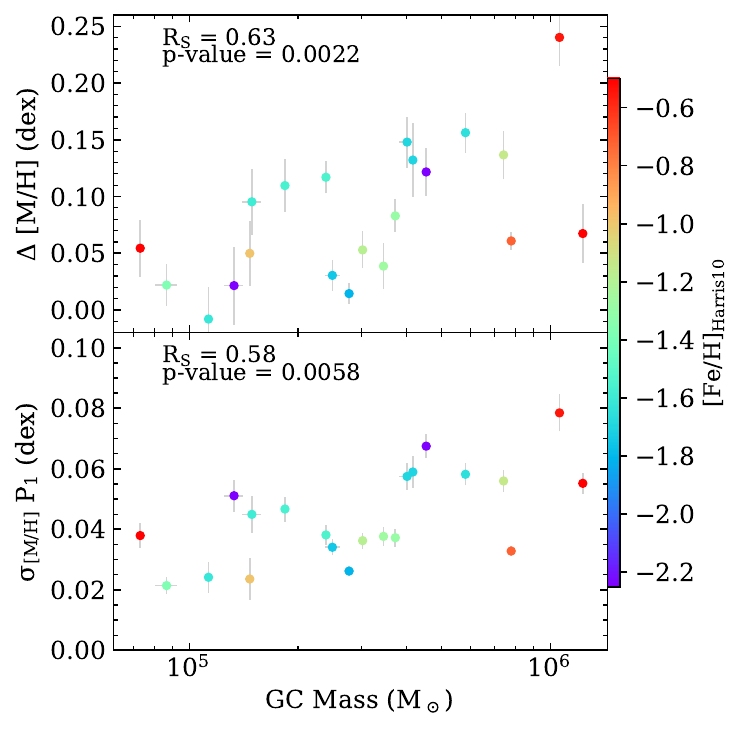}   \includegraphics{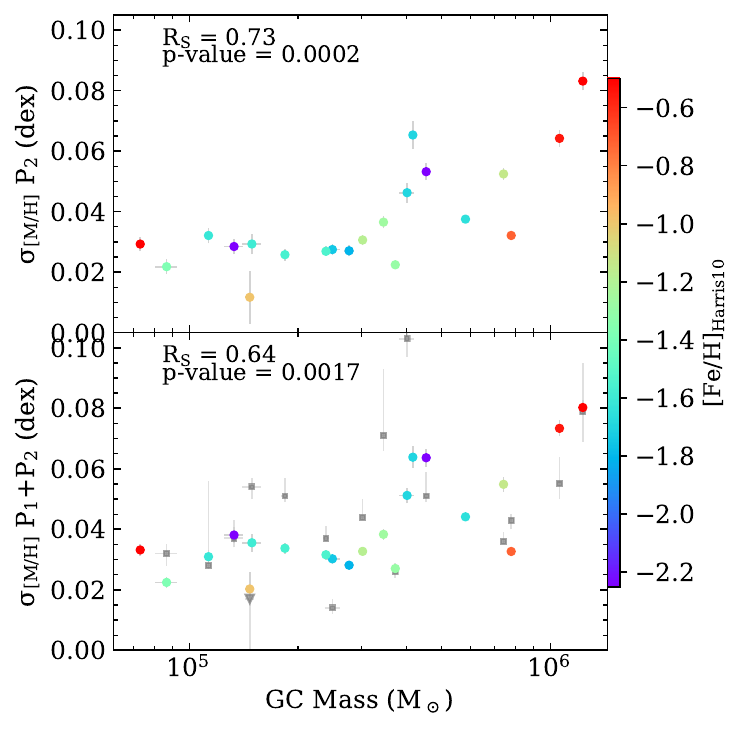}
   }
     \caption{Metallicity spread ($\Delta$\met) and dispersion ($\sigma_{[M/H]}$) versus the mass of the GCs \citep{baumgardt2018}. The metallicity spreads (top left) are for the P1 stars as presented in Sect.~\ref{sec:res:met_spread}. The other panels show the metallicity dispersions for the P1, P2, and P1+P2 stars. 
     Our data points are color-coded by the average metallicity of the cluster (\citeauthor[2010 edition]{harris1996}). The grey squares in the bottom right panel represent the $\sigma_{[Fe/H]}$ taken from the literature compilation of \citet{Bailin19_cat} for the GCs in common with our study. The grey downward triangle indicates the upper limit on $\sigma_{[Fe/H]}$ measured for NGC\,6362.}
     \label{fig:dM_vs_Mass}
\end{figure*}

Concerning the metallicity dispersions for the total P1+P2 populations (see Table~\ref{table_sigmas} and Fig.~\ref{fig:dM_vs_Mass}), we find them to be generally very small, in fact below 0.05 dex in the majority of our clusters, thus fully compatible with the conclusions reached by \citet{carretta2009}, that $\sigma_{\rm [Fe/H]}$ is smaller than 0.05 dex in most of their 19 GCs, and \citet{Bailin19_cat}, that $\sigma_{\rm [Fe/H]}$ is smaller than 0.1 dex with a median value of 0.045 dex for 55 GCs. However, we must keep in mind here that the P3 stars, which are possibly more metal-rich than those of P1 and P2, are excluded from our star samples in the Type II GCs (among our clusters those are NGC\,362, NGC\,1851, NGC\,5286, NGC\,6656, NGC\,6388, and NGC\,7089). So the total, in the sense of considering all RGB stars, metallicity dispersion in these six clusters might be larger. Few studies have measured intrinsic iron, or metallicity, dispersion below 0.05 dex, but \citet{yong2013} did a high-precision differential abundance analysis of RGB stars in NGC\,6752 and estimated an intrinsic iron and metallicity dispersion of $\sim$0.03 dex, which is fully compatible with our value of $\sigma_{\rm [M/H]}$ = 0.032 dex measured for this cluster. Our metallicity dispersion for NGC\,362 of 0.038 dex is also in good agreement with the values obtained by \citet{monty2023} of 0.035 and 0.041 dex (from \ion{Fe}{i} and \ion{Fe}{ii} lines respectively), using the same differential abundance analysis technique (see also \citealt{melendez2009} for a description of the method).

NGC\,6388 is a particular case, with its very large color spread in the chromosome map (\dc=0.8)
we measure a large $\Delta$\met\ of 0.24 dex and the color-metallicity relationship is clearly seen in Fig.~\ref{fig:app2}. It is also the cluster for which we derive the largest dispersion, 0.078 dex, among the P1 stars (0.07 dex for P1$+$P2). We note that the metallicity trends with magnitude and \teff\ discussed in App.~\ref{App_A} are rather mild for this cluster, and by correcting the metallicities accordingly, we only decrease $\Delta$\met\ to 0.20 dex and the total dispersion (P1$+$P2) to 0.068 dex).
This cluster has been classified as Type~II by \citet{mil17} and is suspected to have some intrinsic iron spread \citep{Wallerstein2007,hughes2007}. However, the recent studies by \citet{carretta2022_ngc6388,Carretta_2023}, based on a sample of about 150 bright RGB stars from all populations, concluded that there is no intrinsic iron ($<$~0.04~dex), nor metallicity spread in the cluster, 
thus questioning its classification as a Type~II GC. However, this would leave the significant color spread of the RGB stars in NGC\,6388 unexplained.

\subsection{Comparison with theoretical expectations}\label{sec:diss:model}

The studies of \citet{bailin2018} and \citet{Mckenzie21} simulated the formation of (proto)-GCs and showed that self-enrichment leads to metallicity variations within the resulting GC. Although they use different types of simulations, their models predict a similar outcome in terms of intrinsic metallicity dispersion: an increase in dispersion with the cluster mass for $M>10^5$ \msun. In the top left panel of Fig.~\ref{fig:dM_vs_Mass}, we plotted our resulting $\Delta$\met\ versus the mass of the GCs (from \citealt{baumgardt2018}). We find a positive correlation between the two parameters with a Spearman's rank correlation coefficient $R_{\rm s}=0.63$.
The majority of the GC in our sample have masses in the range of $10^5 - 10^6$ \msun\ where the iron dispersion is expected to slightly increase with the cluster's mass and reach up to $\sim$0.1 dex \citep{bailin2018,Mckenzie21}. Our $\Delta$\met\ reach values that are larger than 0.1 dex, but they cannot be directly compared to the metallicity dispersions ($\sigma$) that are presented in the literature. 
Thus, in the bottom left panel of Fig.~\ref{fig:dM_vs_Mass} we plot the $\sigma_{\rm [M/H]}$ of the P1 stars versus the GC masses. From this dataset, we retrieve a correlation coefficient $R_{\rm S}=0.58$, similar to that obtained from $\Delta$\met, and our dispersion values are between 0.02 and 0.08 dex, in good agreement with the theoretical expectations. 

We also show, in the right panels of Fig~\ref{fig:dM_vs_Mass}, the metallicity dispersion of the P2 and P1+P2 stars versus the GC masses. Because the P2 stars dominate the total sample in most clusters, the behavior is very similar in both cases: the dispersions are essentially constant and at their smallest ($\sigma_{\rm [M/H]} <$ 0.04 dex) until a mass of $\sim 3 \times 10^5$ \msun. At higher masses, the dispersion increases slightly up to $\sim$0.08 dex. This is qualitatively similar to what was measured by \citet{carretta2009} among their sample of 19 GCs using the absolute magnitudes of the clusters as a proxy for the GC mass. We note however that the correlation between iron dispersion and GC mass could not be reproduced by \citet{Meszaros_2020} with their spectra from the SDSS-IV APOGEE-2 survey, but their observed iron dispersions are systematically larger, with an average $\sigma_{\rm [Fe/H]}$ close to 0.1 dex. 

We also compare our dispersion values for P1+P2 with the $\sigma_{[Fe/H]}$ reported in the literature compilation of \citet{Bailin19_cat} (updated in \citealt{Bailin2022}) for the GCs in our study (17 are in common, shown with grey symbols in Fig.~\ref{fig:dM_vs_Mass}). Both samples show a similar range of dispersion although the correlation between the 17  \citet{Bailin19_cat} data points and the GC mass is weaker ($R_{\rm S}=0.44$). It is worth specifying here that \citet{Bailin19_cat} considered iron measurements from RGB stars regardless of their population, so it presumably includes a mix of P1 and P2 stars but also stars belonging to the red RGB in the case of type II clusters. The ratio of stars from each population included in their spectroscopic samples varies from one cluster to another. 

In their photometric study of iron variations within P1 stars, \citet{Legnardi22} observed a negative correlation between $\Delta$[M/H] and the average GCs metallicity. Within our sample, we do not have a correlation ($R_S <$ 0.13) between neither the metallicity spread ($\Delta$\met) nor any of the dispersions (P1, P2, P1+P2) and the average metallicity of the GC. We note that there was also no such correlation among the observed $\sigma_{[Fe/H]}$ of \citet{Bailin19_cat} (see also \citealt{Bailin2022}) and no global correlation is expected from their theoretical models either. There is a relationship expected between metallicity and dispersion, but it is linked to the initial metallicity of the model (see Fig.~1 of \citealt{Bailin2022}). It essentially predicts a maximum possible dispersion that decreases with increasing metallicity.

\section{Conclusion}\label{sec:concl}

We used the metallicities of RGB stars in 21 GCs to quantify the metallicity spread among the primordial stellar population of each cluster and characterize its intrinsic metallicity dispersion. The stellar metallicities are derived from fitting the MUSE spectra with model atmospheres and, for each star, we averaged the results from at least three individual spectra with $S/N$ $>$ 20. The exceptions are NGC\,6681 and NGC\,6362 for which we have only two and one available measurement per star, respectively.

We characterized the relationship between the stellar metallicities [M/H] and the pseudo-color \cmapx, which is the position of the star on the x-axis of the chromosome map. In almost all clusters, we observe a clear trend of increasing metallicity with increasing values of \cmapx\ among the P1 stars.  
From this relationship, we measure metallicity differences that mostly range from 0.03 to 0.15 dex. As expected, the metallicity variations are generally found to be larger in the GCs showing a larger \cmapx\ color extension of their P1 stars. In this case, NGC\,6388 stands out with, by far, the largest color spread among its P1 stars,
resulting in a metallicity spread of 0.24 dex. Our metallicity spreads are in good agreement with those estimated in various literature works based on spectroscopy, but we find them to be, on average, systematically lower than those measured from photometry by \citet{Legnardi22}. 
We also find that the metallicity spread within the P1 stars correlates with the mass of the GCs.

In addition, we investigated the intrinsic metallicity dispersion ($\sigma_{\rm [M/H]}$) of the P1 stars, the P2 stars, and the combination of both (P1+P2). Our dispersion values are all relatively small ($<$0.08 dex) and comparable, if not smaller, to values presented in literature work using high-resolution spectroscopy \citep{carretta2009,yong2013,Meszaros_2020,monty2023}.
For both the P1 and P2 stars we find a correlation between the intrinsic dispersion and the mass of the GC. It is particularly striking when looking at the metallicity dispersion of the P2 stars where the GCs less massive than $\sim 3 \times 10^5$ \msun\ all have similarly low dispersion ($<$0.04 dex) but for the more massive GCs, the dispersion increases with the cluster mass. Interestingly, we find that for all but one cluster, the metallicity dispersion of the P2 stars is lower, or equal to, that of the P1 stars. This is in line with the results of \citet{Legnardi22} who concluded, from photometry, that in NGC\,6362 and NGC\,6838, the metallicity spread among the P2 stars is lower than in the P1 stars. Our results show that this is a general tendency among the massive GCs of the Milky Way.

Even though the resolution of the MUSE spectra is too low to measure iron abundances from individual iron lines, we can achieve high precision in measuring small metallicity variation and dispersion thanks to the sheer amount of spectra accumulated over the last decade as part of the MUSE GTO GC survey. In future work, we will explore in further detail the metallicities of the anomalous population (i.e., the red-RGB) of Type-II GCs and some peculiarities among the P2 stars of selected GCs like NGC\,104 and NGC\,7078. 

\section{Data availability}\label{data_availability}

Table 1 is only available at the CDS via anonymous ftp to
cdsarc.u-strasbg.fr (130.79.128.5) or via http://cdsarc.u-strasbg.fr/viz-bin/cat/J/A+A/XXX/zzz

\begin{acknowledgements}
We thank N. Bastian, M. Salaris, and S. Martens for useful comments and discussion.
M.L. acknowledges funding from the Deutsche Forschungsgemeinschaft (grant LA 4383/4-1) and from the German Ministry for Education and Science (BMBF Verbundforschung) through grants 05A14MGA, 05A17MGA, 05A14BAC, 05A17BAA, and 05A20MGA. 
S.K. gratefully acknowledges funding from UKRI in the form of a Future Leaders Fellowship (grant no. MR/T022868/1).
S.M. was supported by a Gliese Fellowship at the Zentrum f\"ur Astronomie, University of Heidelberg, Germany.
This research has made use of the Astrophysics Data System, funded by NASA under Cooperative Agreement 80NSSC21M00561 and the \textsc{python} packages \textit{pandas} \citep{reback2020pandas,mckinney-proc-scipy-2010}, \textsc{matplotlib} \citep{Hunter:2007}, and statsmodels \citep{seabold2010statsmodels}.
\end{acknowledgements}


\bibliographystyle{aa}


%
%

\begin{appendix}

\section{On the metallicity$-$magnitude trend}\label{App_A} 
Upon inspection of our data, we found that our metallicities ([M/H]) are not constant along the RGB: they increase with increasing luminosity. Given the relationship between luminosity, \teff\ and log~$g$ for RGB stars, this also means that the metallicities decrease with increasing log~$g$ and \teff. Similar issues were reported from the MUSE spectra of NGC\,6397 obtained as part of the instrument commissioning \citep{husser2016,baratella2022}. From stellar evolution models including atomic diffusion, it is known that stars near the main-sequence turnoff display a lower metallicity due to the onset of diffusion processes as convection in the outer layers of the star ceases. Among other radiative transport processes, gravitational settling causes the heavier elements to sink at the bottom of the atmospheres, thus reducing the photospheric metallicity \citep{VandenBerg2002,Nordlander2012}. As the stars evolve on the subgiant branch, convection re-appears at the surface and the metallicity gradually returns to its original value as the heavy elements are brought back to the surface. According to stellar evolution calculations and isochrones taking into account atomic diffusion (e.g.,\citealt{gruyters2014,hidalgo2018,Choi2016}), the original metallicity is still not fully recovered at the bottom of the RGB, such that a difference in photospheric metallicity of $\approx$0.02-0.03 dex can be expected between the bottom and the middle part of the RGB. We note that this effect is expected to be larger at lower metallicities.
The effect of atomic diffusion is nicely illustrated in Fig. 10 of \citet{nitschai23} where metallicity changes are seen in the measured metallicities from MUSE spectra in $\omega$~Centauri. Metallicity differences predicted from theoretical isochrones are also shown in that same figure.

\begin{figure}
\resizebox{\hsize}{!}{
   \includegraphics{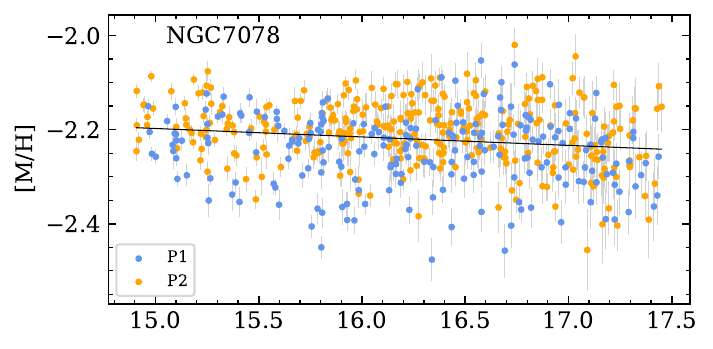}}\vspace{1pt}
\resizebox{\hsize}{!}{   
   \includegraphics{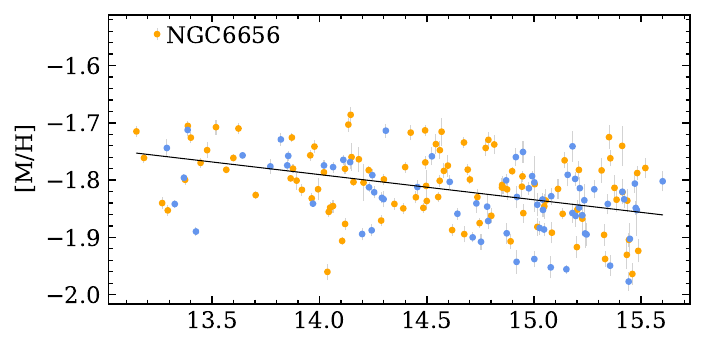}
   }
\resizebox{\hsize}{!}{
   \includegraphics{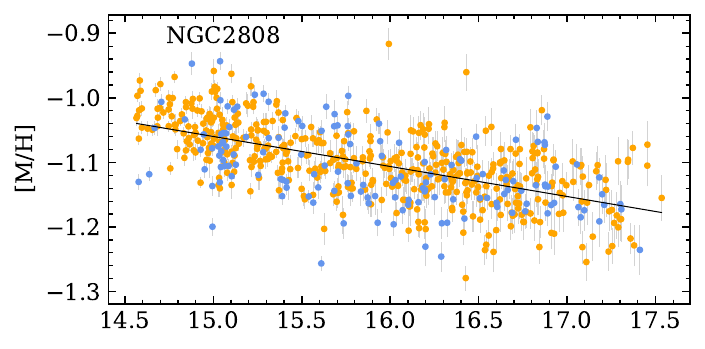}}\vspace{1pt}
   \resizebox{\hsize}{!}{   
   \includegraphics{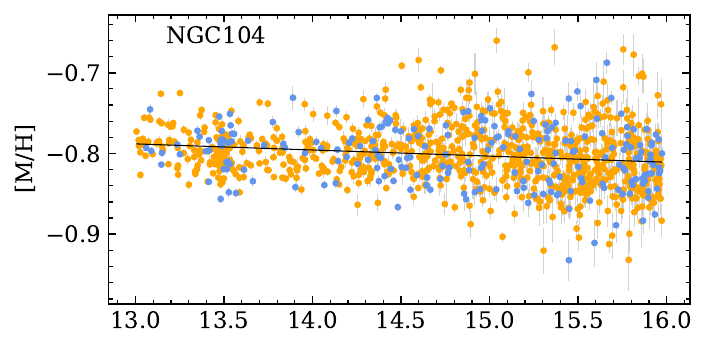}
   }   
\resizebox{\hsize}{!}{   
   \includegraphics{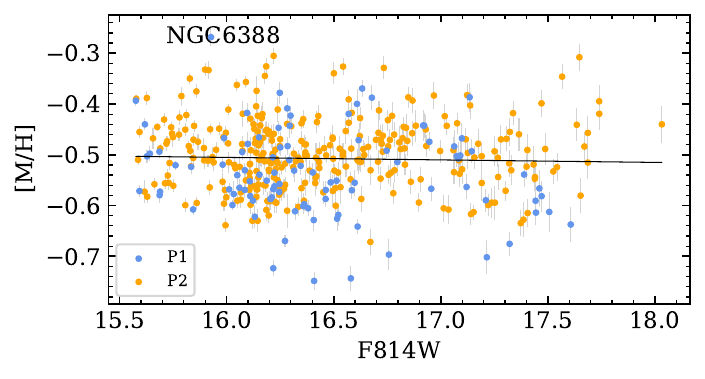}
   }   
     \caption{ The metallicity$-$magnitude trend for five clusters spanning the metallicity range between [M/H]=$-$2.2 and $-$0.5 dex. The P1 and P2 stars are indicated by blue and orange dots respectively.
The black line indicates the best fit linear relationship for all stars.}\label{fig:app_tefftrend}
\end{figure}

\begin{figure}
   \includegraphics[width=0.5\textwidth]{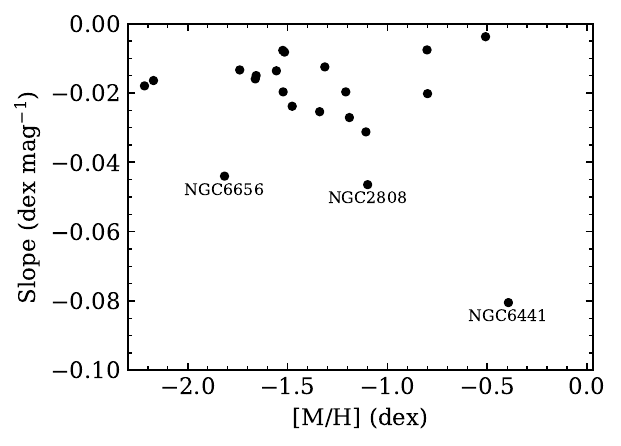}
     \caption{ Slope of the \met-$F814W$ relation versus the average [M/H] of the stars in our clusters.}
     \label{fig:App_slopes}
\end{figure}

The gradient seen in our data, in terms of \met\ versus magnitude ($F814W$), is more or less important depending on the cluster. We show it for five GCs in Fig.~\ref{fig:app_tefftrend}. In some clusters, like NGC\,104, NGC\,7089, and NGC\,6752, the effect is mild (less than 0.03 dex difference between the faintest and brightest stars in our sample) and could be explained by atomic diffusion. But in some other cases, like in NGC\,6441 and NGC\,2808, the effect is much stronger than expected (more than 0.1 dex). 
We investigated our data and analysis method, but we could not find an explanation for this large \met$-$magnitude trend. We do not see a correlation with the metallicity of the clusters, as is expected from diffusion. We show in Fig. \ref{fig:App_slopes} the slope derived versus the average metallicity of the cluster. The three clusters with a slope larger than 0.04 are NGC\,6656, NGC\,2808, and NGC\,6441.

To verify if this affects significantly our results, we computed "corrected" [M/H] values by removing the trend in the [M/H]$-F814W$ plane. For that, we fitted a linear relationship between [M/H] and $F814W$ for the P1 and P2 stars, individually. Then, we subtracted to the average metallicity the residual between the measured value of [M/H] and the predicted value from the linear relationship. 
Then we redid our analyses using the corrected metallicity for each star. 

\begin{figure}
   \includegraphics[width=0.5\textwidth]{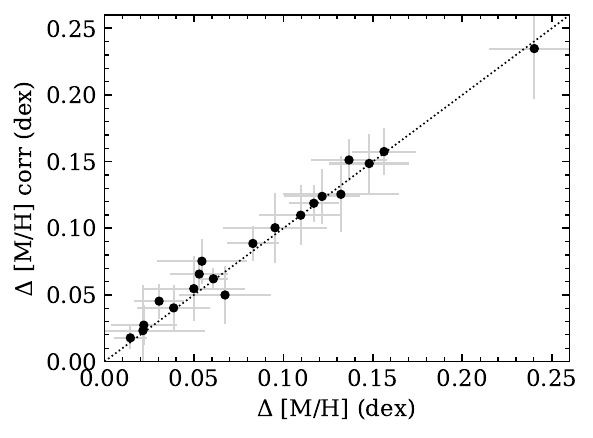}
     \caption{Comparison between the metallicity spreads $\Delta$\met\ 
     }
     \label{fig:App_dM_vs_corr}
\end{figure}

\begin{figure}
   \includegraphics[width=0.5\textwidth]{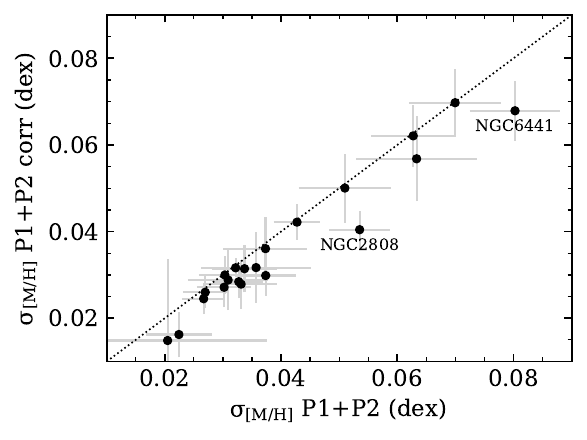}
     \caption{Comparison between the metallicity dispersions  $\sigma_{\rm [M/H]}$ 
     obtained with the original and corrected metallicities.}
     \label{fig:App_sigma_vs_corr}
\end{figure}

\begin{figure}
   \includegraphics[width=0.5\textwidth]{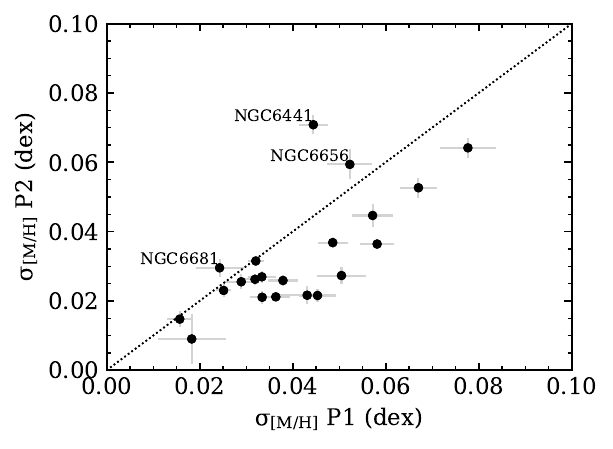}
     \caption{Same as Fig.~\ref{fig:sigmaP1P2} but for the dispersions computed with the corrected metallicities.}
     \label{fig:App_sigmaP1P2corr}
\end{figure}

The metallicity spreads ($\Delta$\met\ corr) obtained with the corrected metallicities are consistent, within their uncertainties, with the results presented in the main text (see Fig.~\ref{fig:App_dM_vs_corr}). In the case of the metallicity dispersions, they decrease slightly but in most clusters, the decrease in dispersion is within the 3$\sigma$ uncertainties (see Fig.~\ref{fig:App_sigma_vs_corr}). The two exceptions are NGC\,6441 and NGC\,2808, those are the two GCs with the strongest \met$-$ magnitude gradient. The $\sigma_{\rm [M/H]}$ obtained with the corrected metallicities are significantly lower in these two clusters. For NGC\,2808 $\sigma_{\rm [M/H]}$ decreases from 0.055 to 0.041 dex and for NGC\,6441 it decreases from 0.080 to 0.068 dex.
When using the corrected metallicities, the P1 stars are still showing a larger dispersion than the P2 stars, except for the same three GCs (see Fig.~\ref{fig:App_sigmaP1P2corr}). Finally, the correlations between the GC masses and the metallicity spread ($\Delta$\met) and dispersion essentially remain the same whether we use the original or corrected metallicities. 
As a final remark, we mention that we also did this exercise of correcting individual metallicities to remove the \met $-$ \teff\ gradient and the results are not significantly different than what is presented here except for the $\sigma_{\rm [M/H]}$ of NGC\,6441 that decreases down to 0.062 dex.



\clearpage

\onecolumn
\section{Additional Figures}\label{App_B}


\begin{figure}[hb!]
\centering
\resizebox{\hsize}{!}{
   \includegraphics{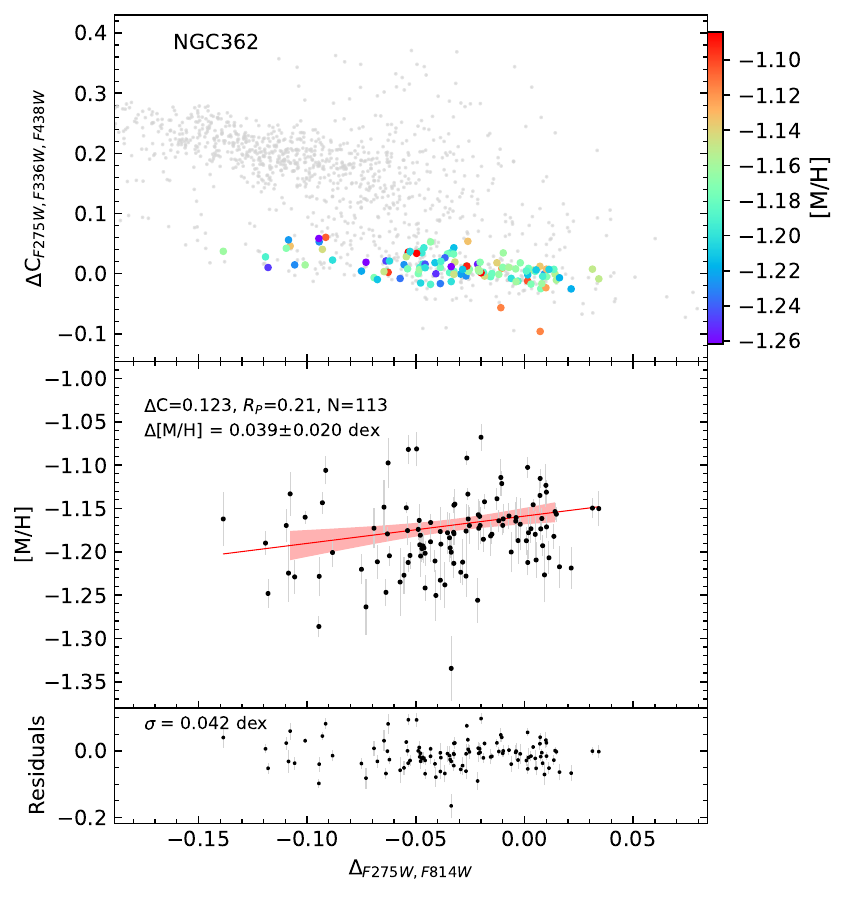}
   \includegraphics{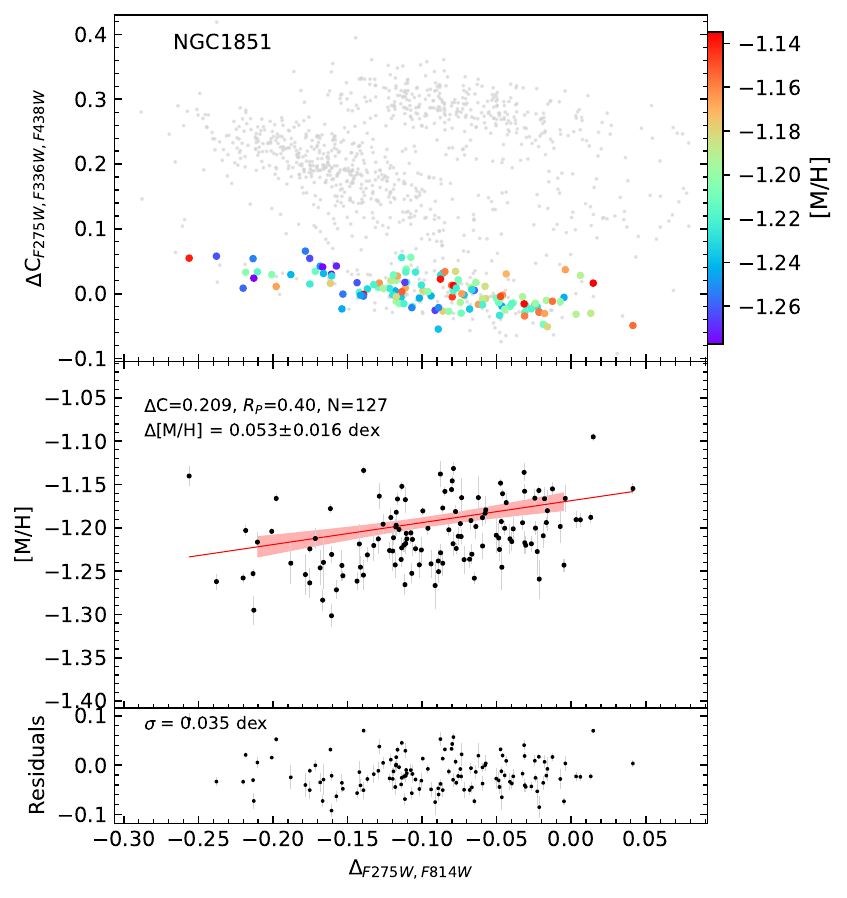}}\vspace{1pt}
\resizebox{\hsize}{!}{   
   \includegraphics{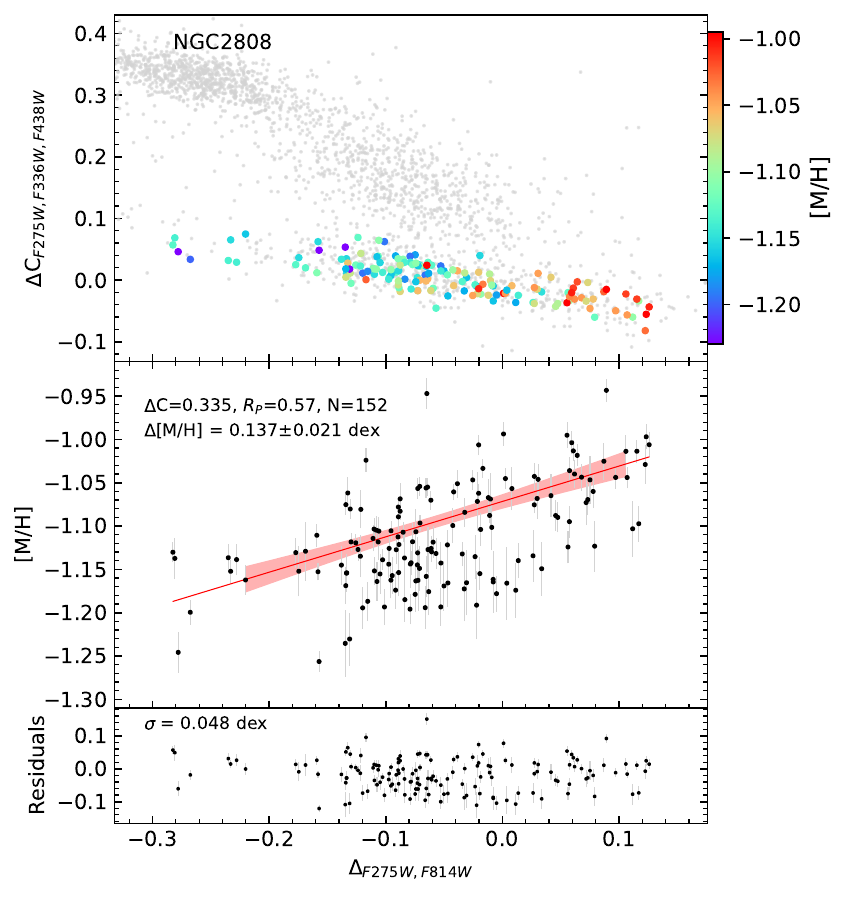}
   \includegraphics{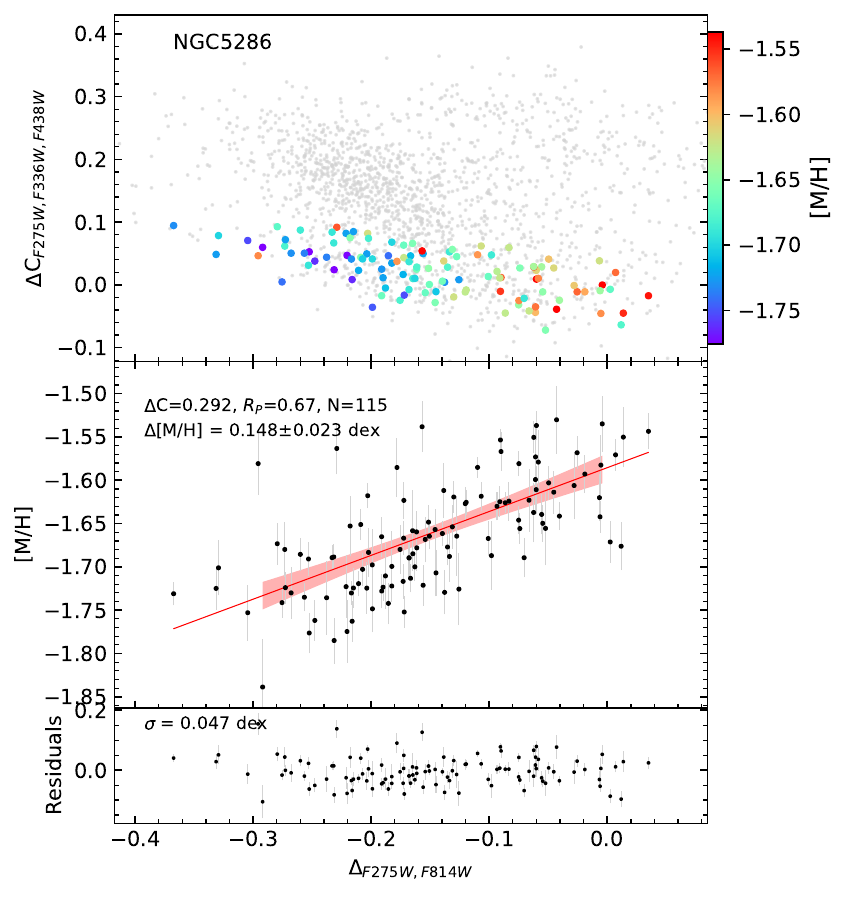}   
   }
     \caption{Same as Fig.~\ref{fig:plots} but for NGC\,362, NGC\,1851, NGC\,2808, and NGC\,5286. Metallicity \met\ versus $\Delta_{\rm F275W,F814W}$ pseudo-color relationship. On the top panels, we show the chromosome maps of the clusters with the P1 stars within our sample color-coded by their metallicity. The middle panels show the metallicity of each star, with the relationship derived from the WLS regression (red line) and the 95\% confidence interval (red shaded area) over the color range \dc. We also indicate the number of stars ($N$), the Pearson correlation coefficient ($R_{\rm P}$), and the resulting metallicity variation $\Delta$\met. The bottom panels show the residuals as in observed$-$predicted and we indicate the standard deviation ($\sigma$) of the residuals. 
     }\label{fig:app1}
\end{figure}

\begin{figure}
\resizebox{\hsize}{!}{
   \includegraphics{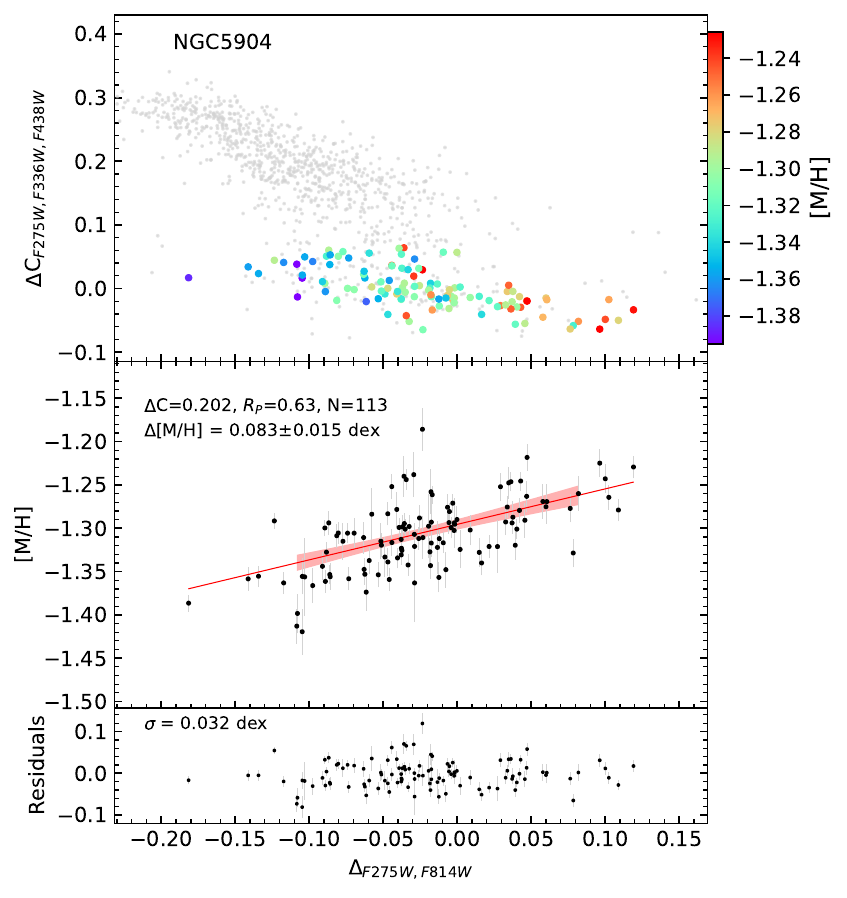}
   \includegraphics{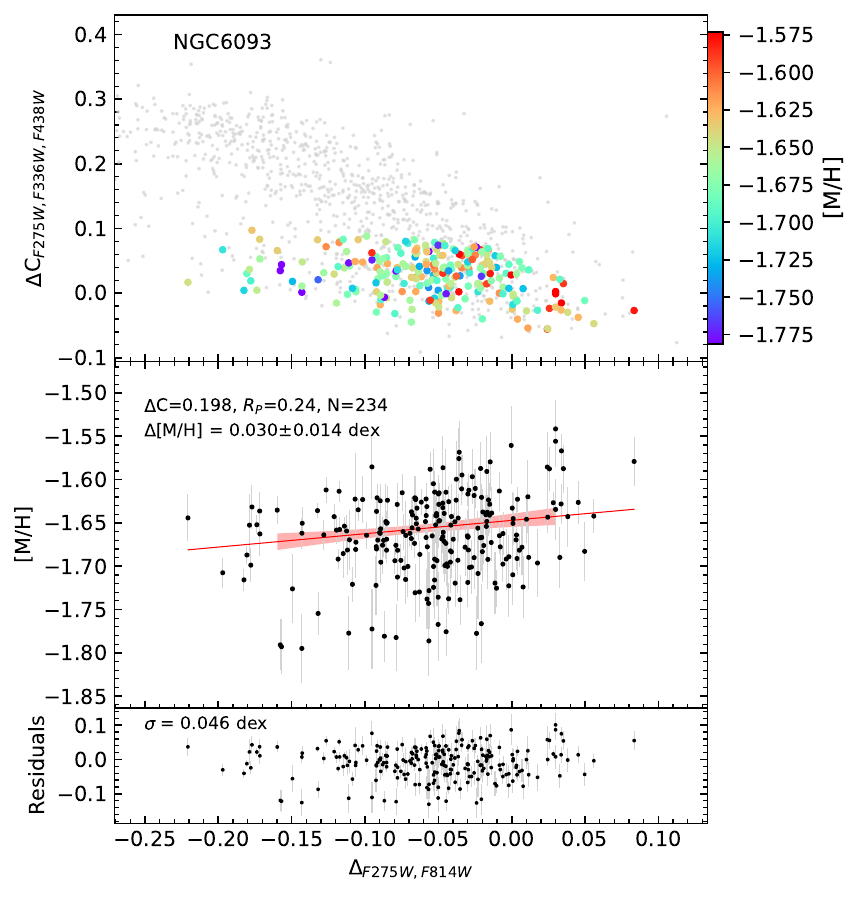}}\vspace{1pt}
\resizebox{\hsize}{!}{   
   \includegraphics{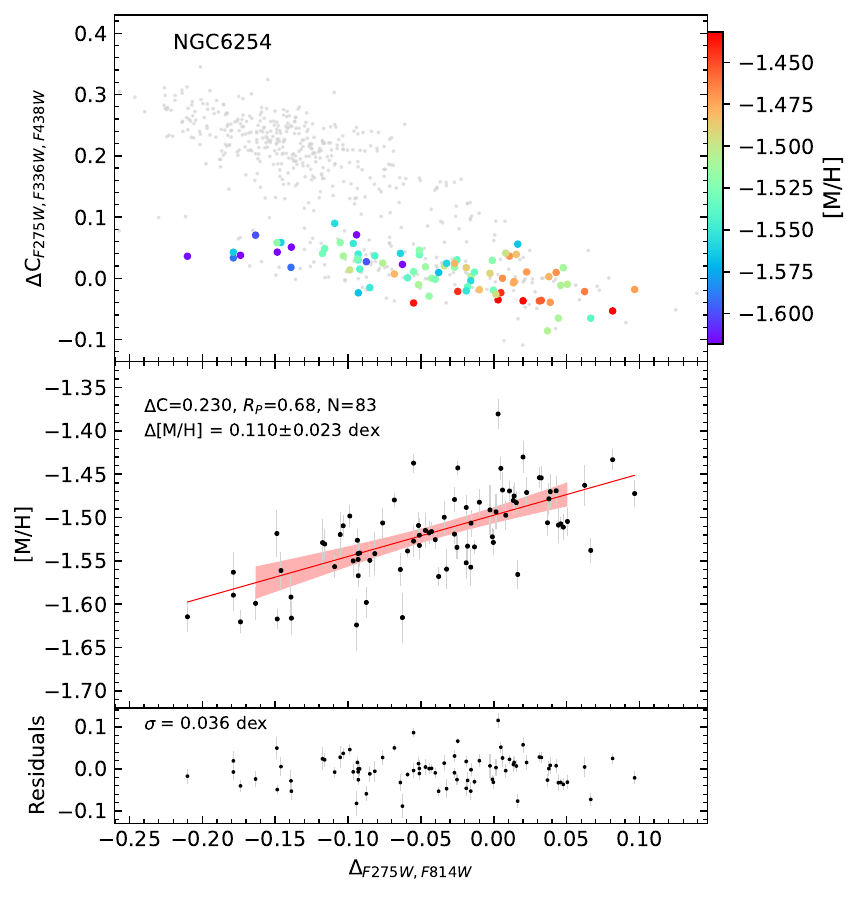}
   \includegraphics{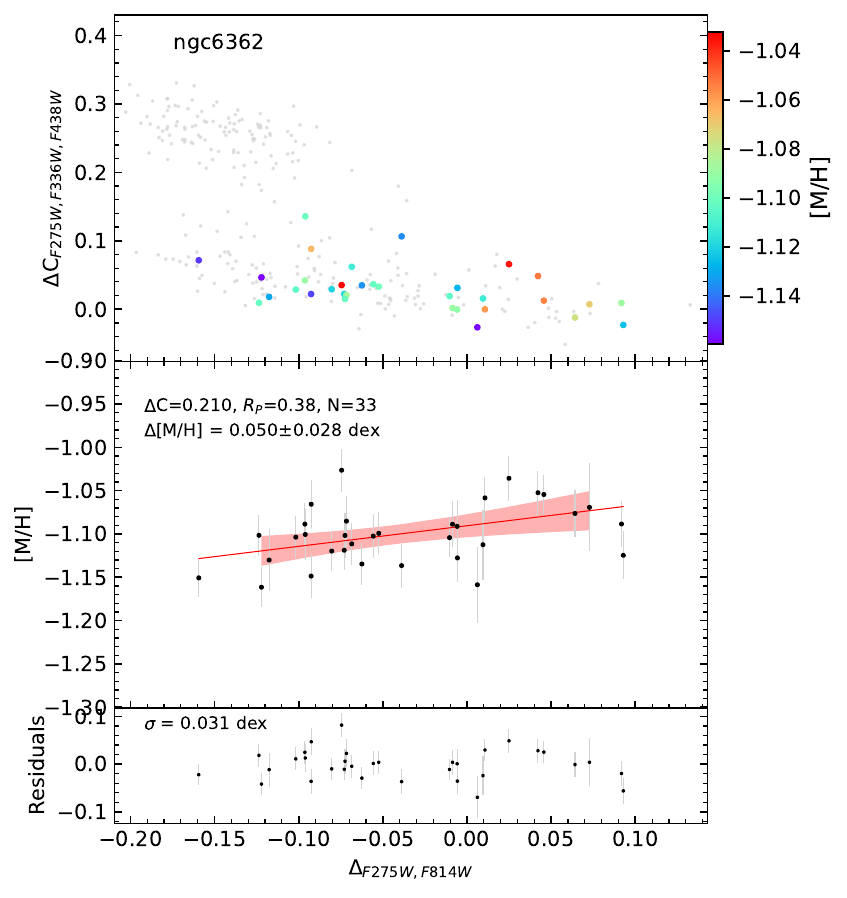}   
   }
     \caption{Same as Fig.~\ref{fig:app1} for NGC\,5904, NGC\,6093, NGC\,6254, and NGC\,6362}\label{fig:app2}
\end{figure}

\begin{figure}
\resizebox{\hsize}{!}{
   \includegraphics{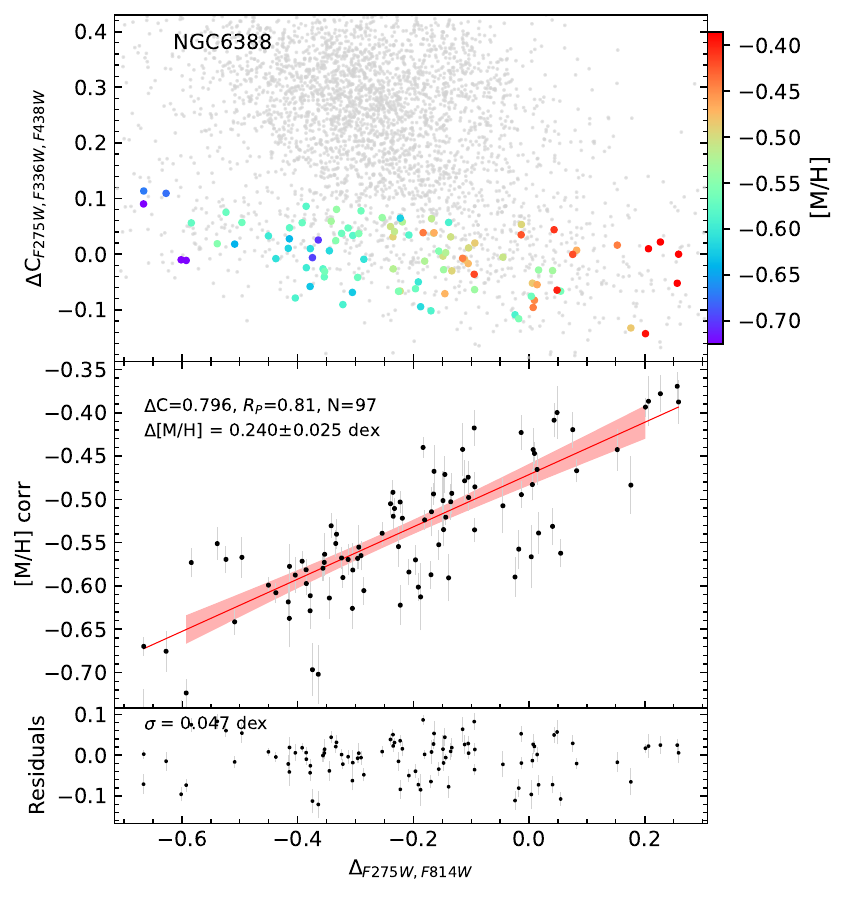}
   \includegraphics{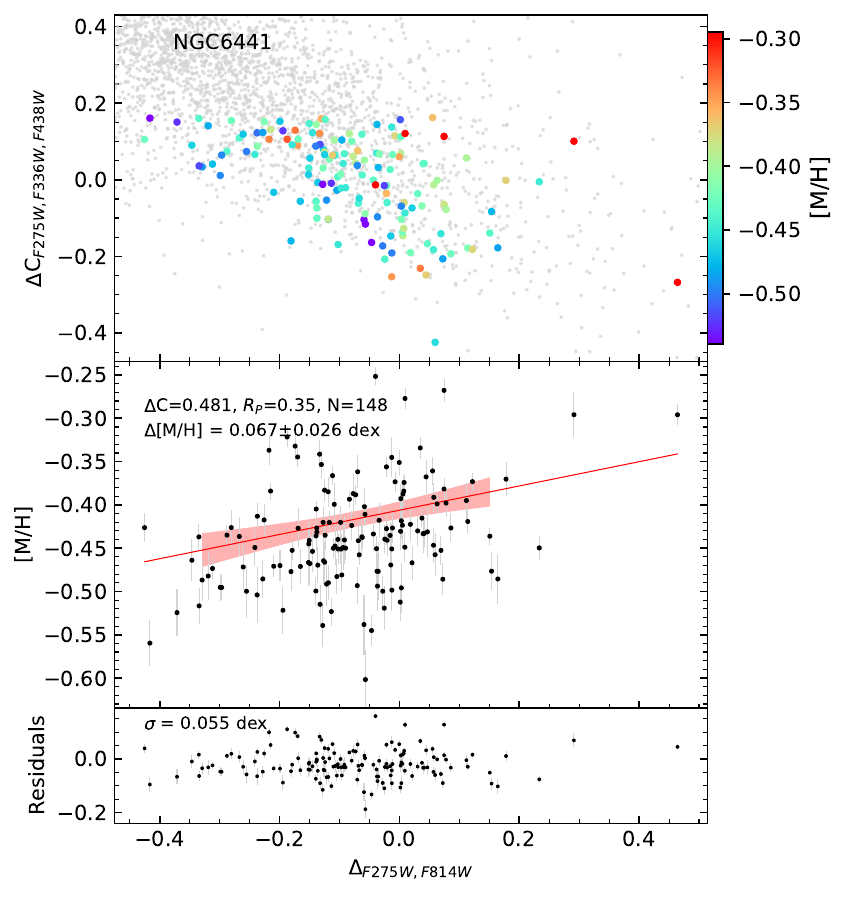}}\vspace{1pt}
\resizebox{\hsize}{!}{   
   \includegraphics{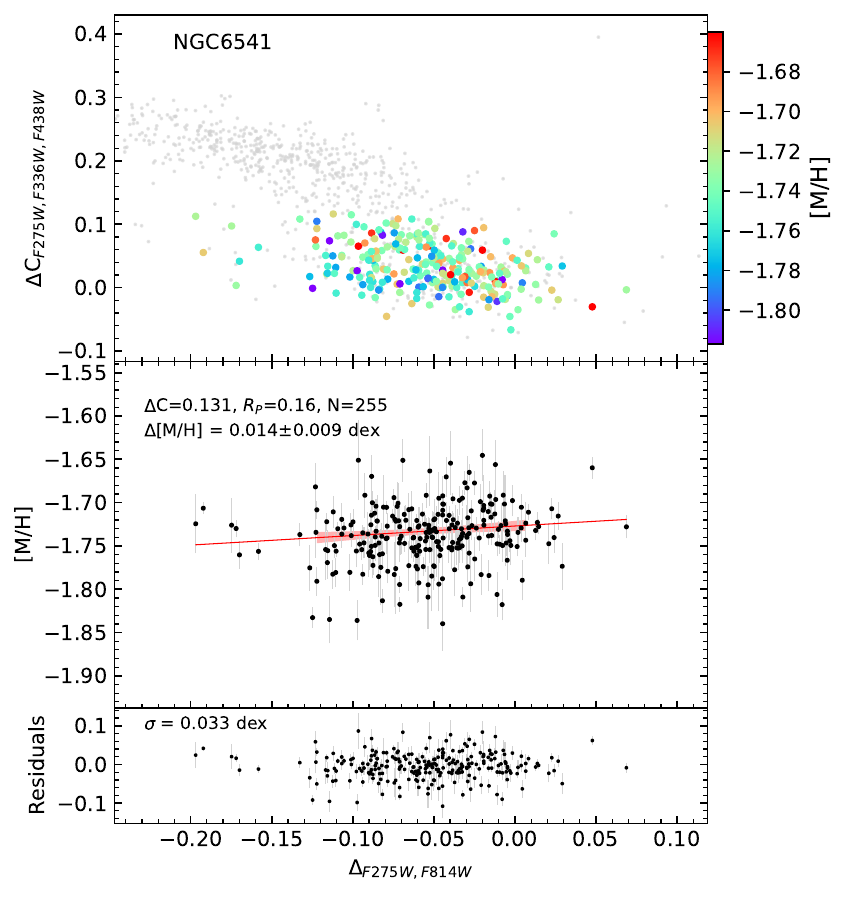}
   \includegraphics{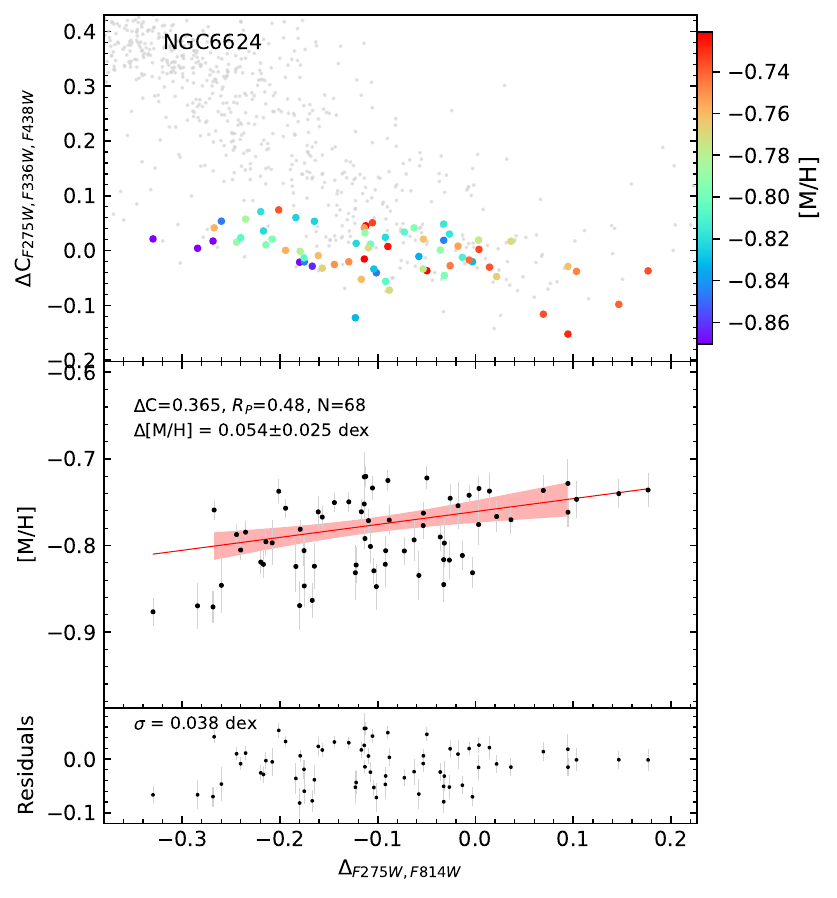}   
   }
     \caption{Same as Fig.~\ref{fig:app1} for NGC\,6388, NGC\,6441, NGC\,6541, and NGC\,6624}\label{fig:app3}
\end{figure}

\begin{figure}
\resizebox{\hsize}{!}{
   \includegraphics{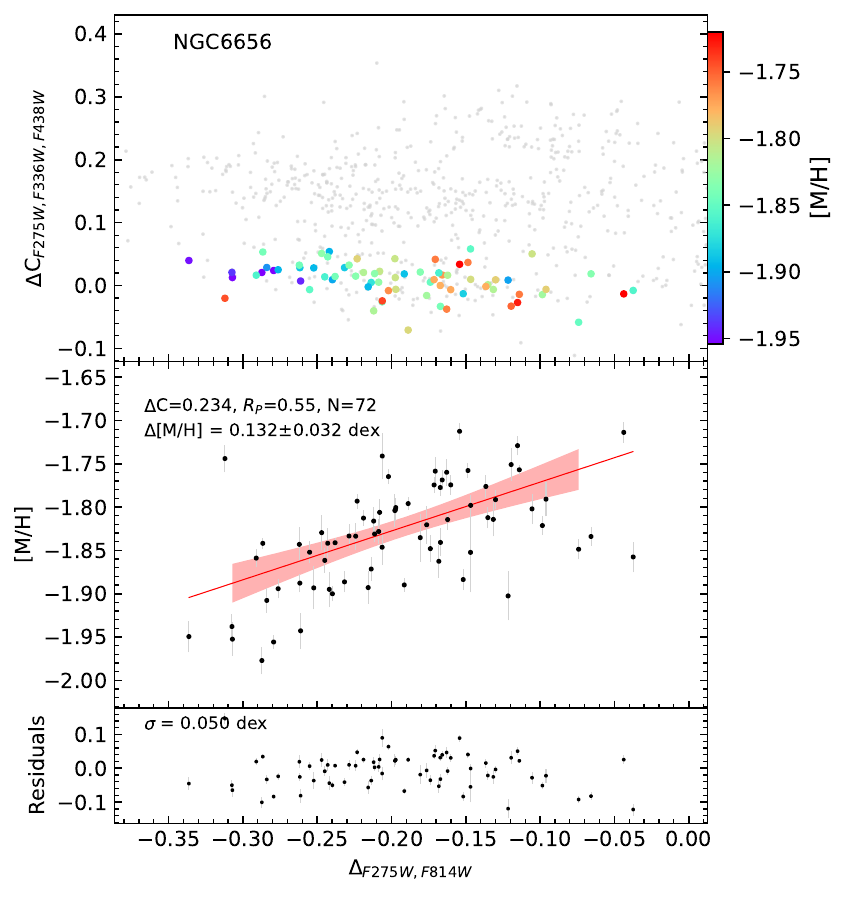}
   \includegraphics{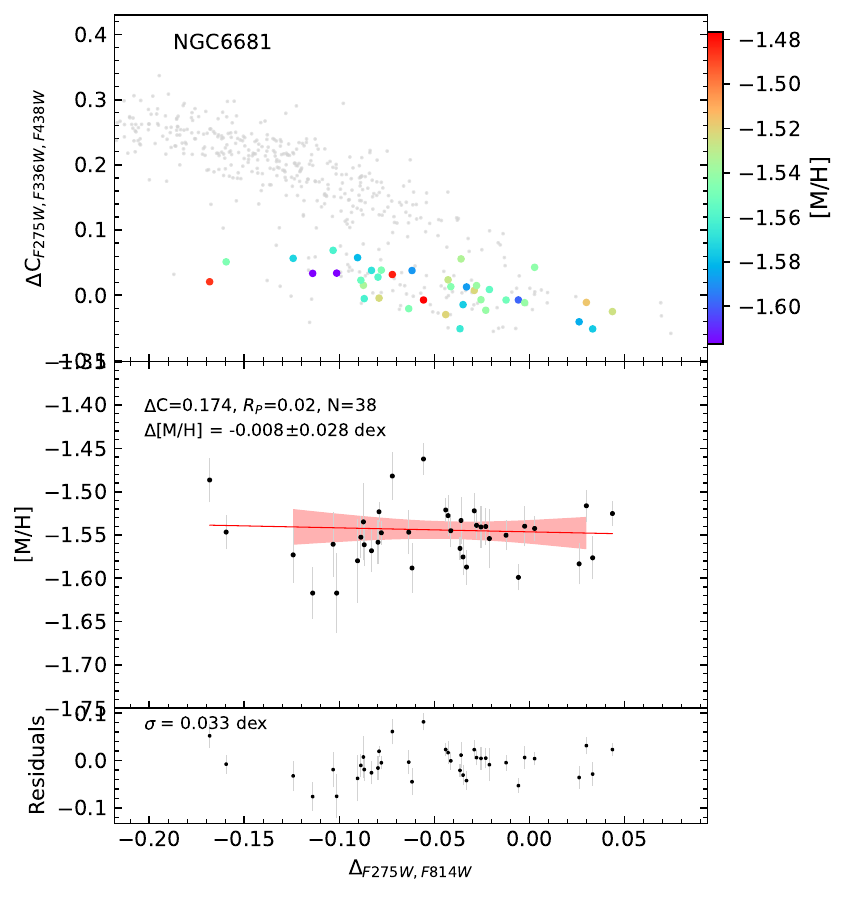}}\vspace{1pt}
\resizebox{\hsize}{!}{   
   \includegraphics{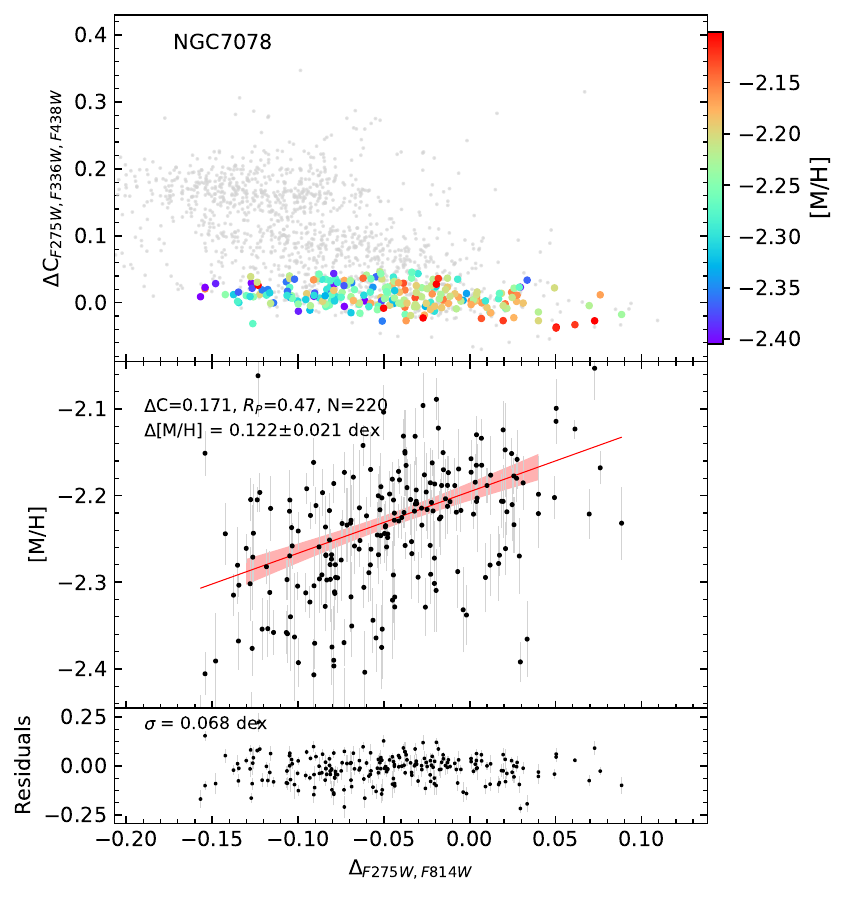}
   \includegraphics{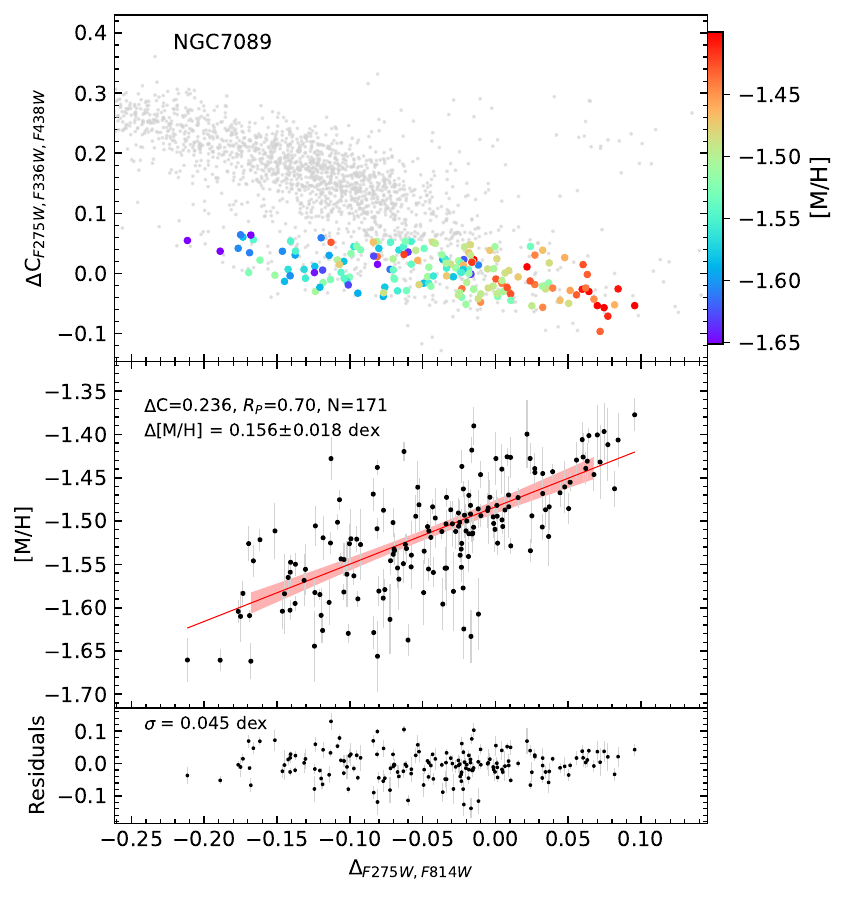}   
   }
     \caption{Same as Fig.~\ref{fig:app1} for NGC\,6656, NGC\,6681, NGC\,7078, and NGC\,7089}\label{fig:app4}
\end{figure}

\begin{figure}
   \includegraphics[width=0.5\textwidth]{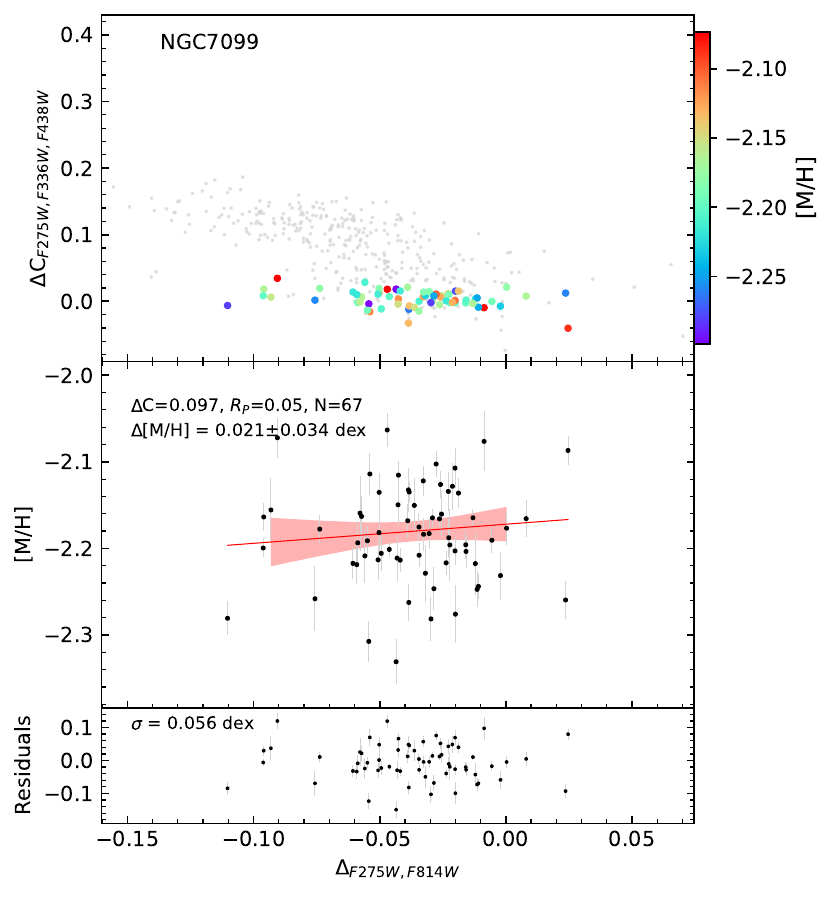}
     \caption{Same as Fig.~\ref{fig:app1} but for NGC\,7099}
     \label{fig:app5}
\end{figure} 


\begin{figure}
\resizebox{\hsize}{!}{
   \includegraphics{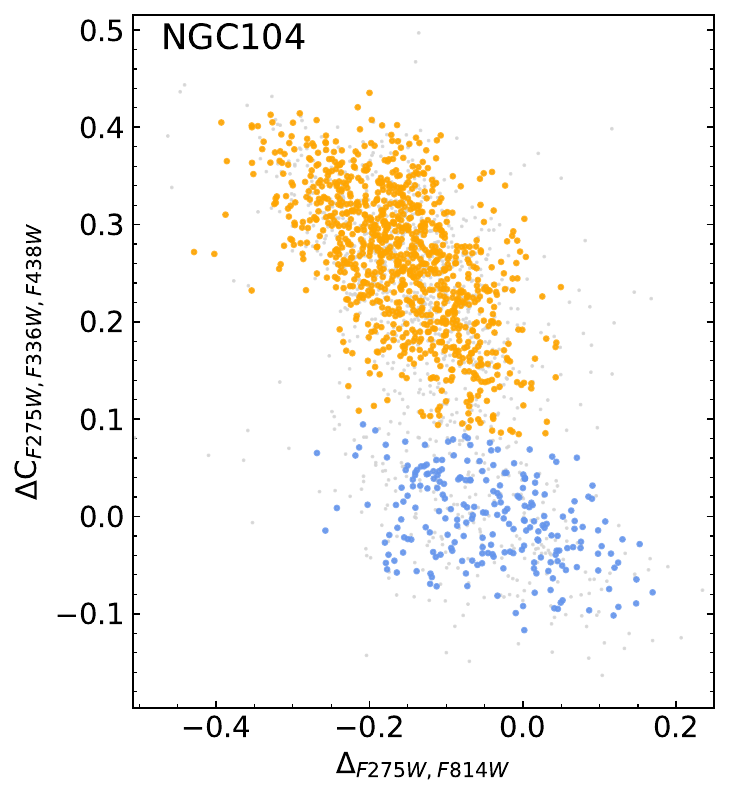}
   \includegraphics{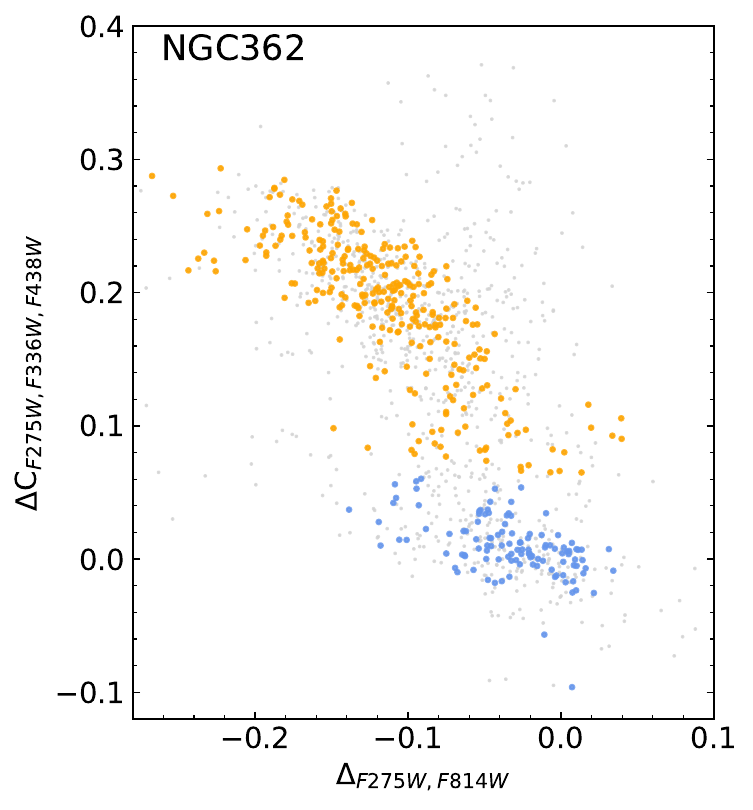}
   \includegraphics{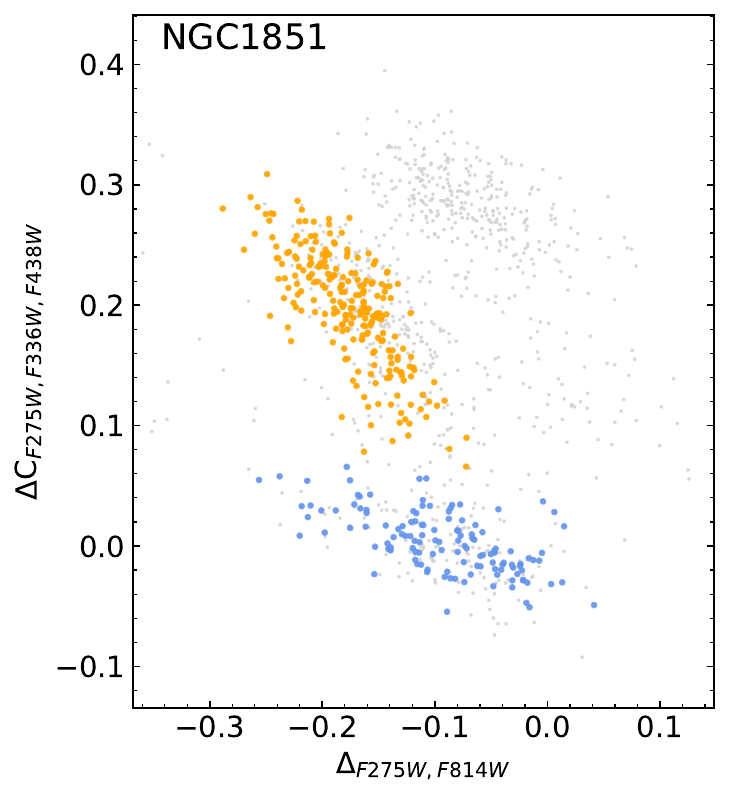}
   }\vspace{1pt}
     \caption{Chromosome maps for NGC\,104, NGC\,362, and NGC\,1851. The stars included as P1 and P2 stars in our analyses are indicated with blue and orange dots respectively. The small grey dots represent all the stars in our chromosome maps.}\label{fig:app_cmap1}
\end{figure}



\begin{figure}
\resizebox{\hsize}{!}{
     \includegraphics{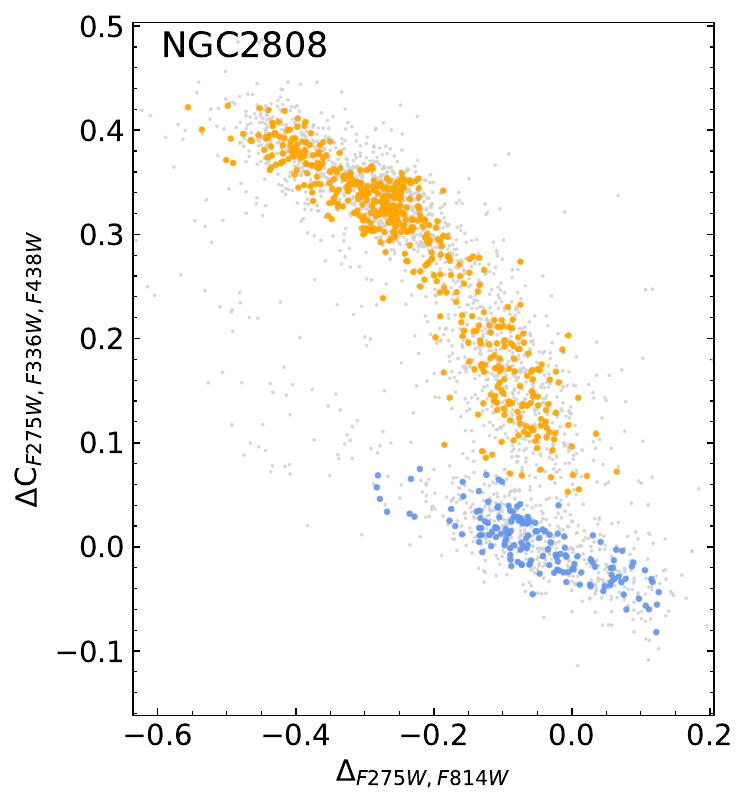}
   \includegraphics{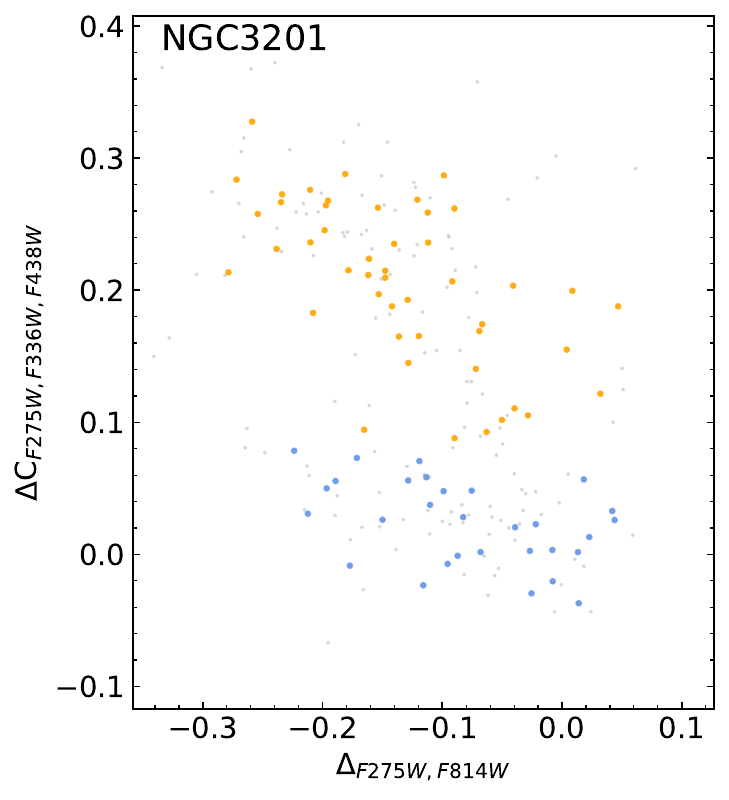}
   \includegraphics{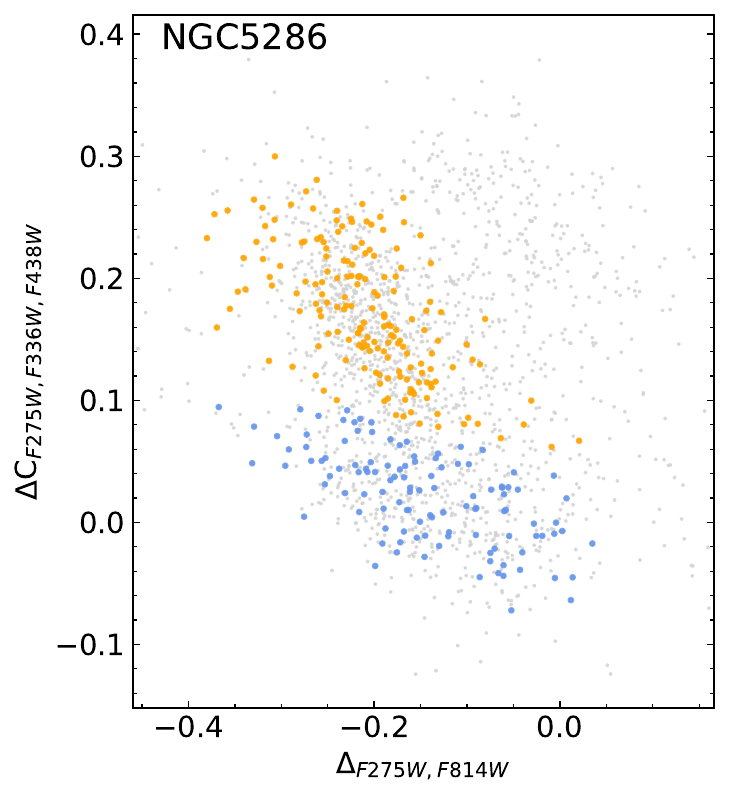}
   }\vspace{1pt}
\resizebox{\hsize}{!}{   
   \includegraphics{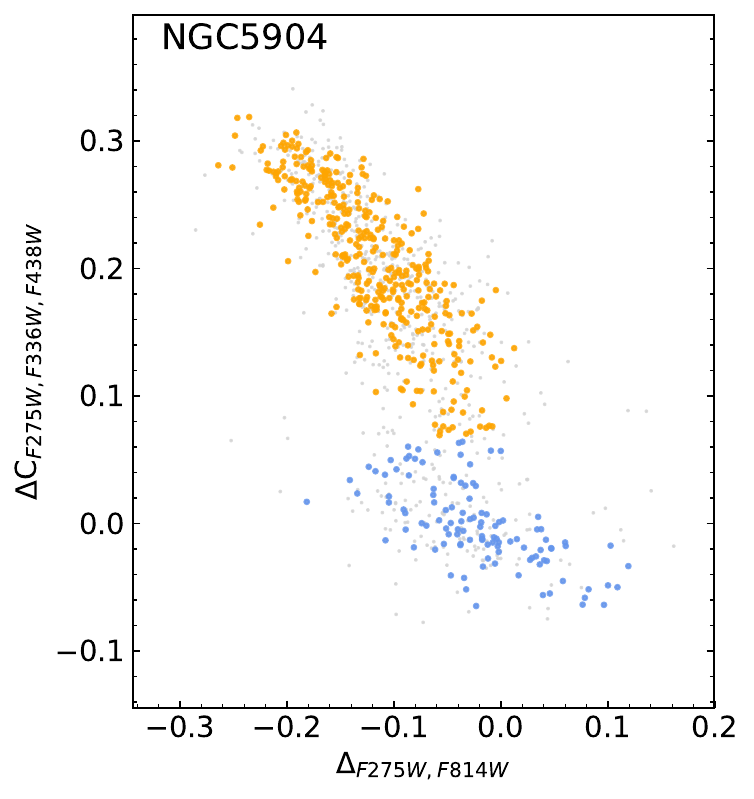}
   \includegraphics{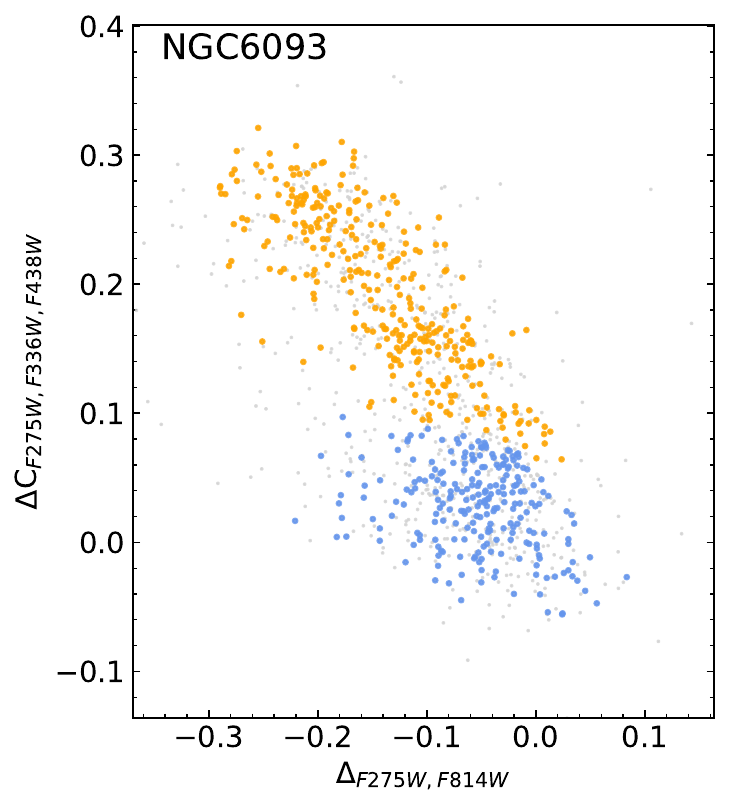}
   \includegraphics{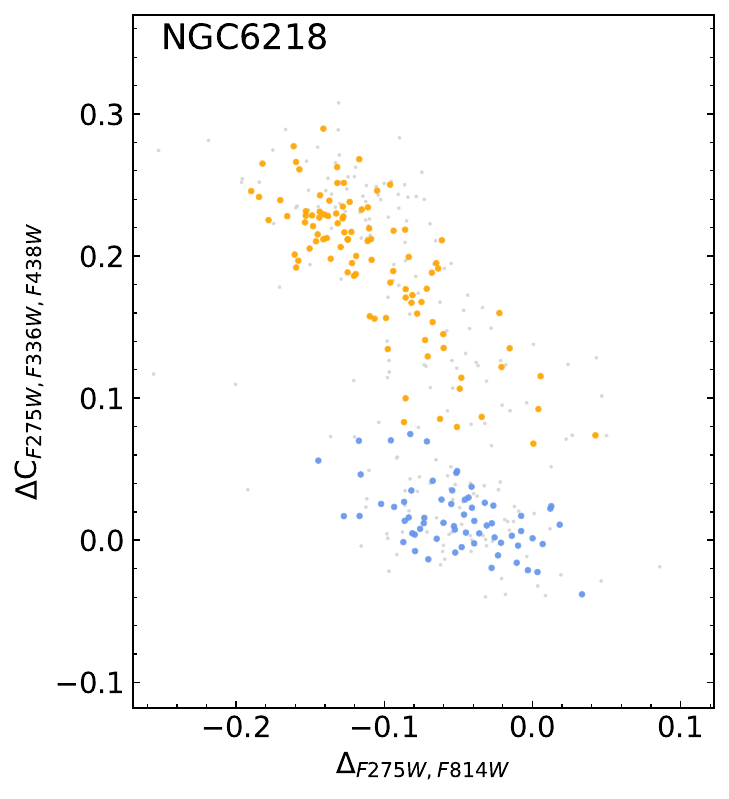}
   }\vspace{1pt}
\resizebox{\hsize}{!}{
   \includegraphics{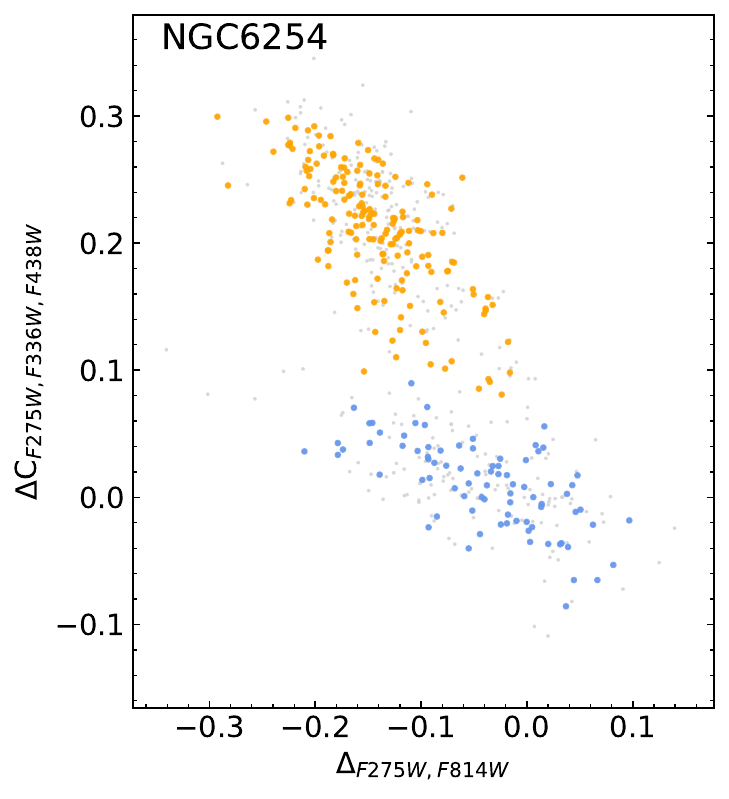}
   \includegraphics{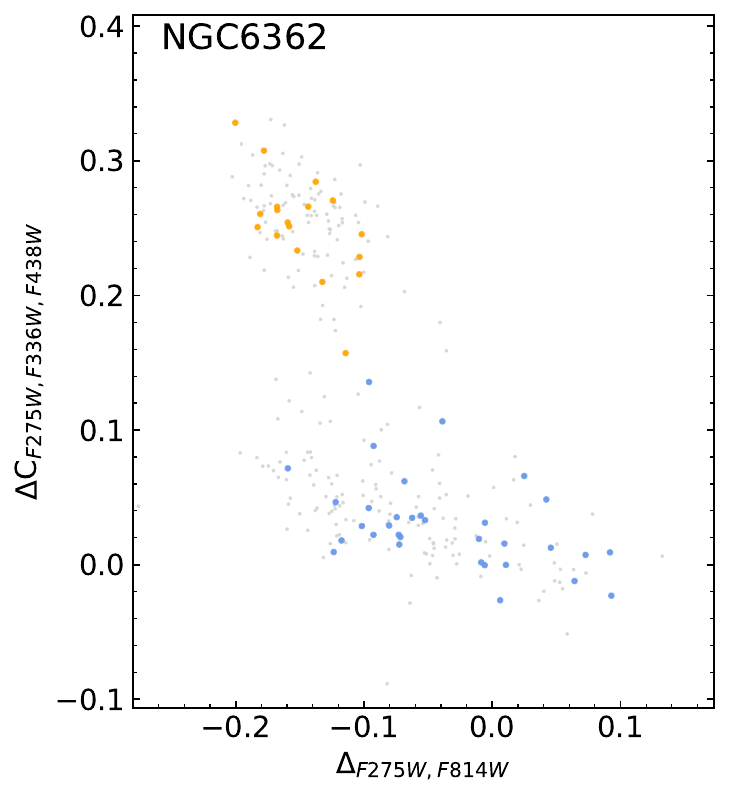}
   \includegraphics{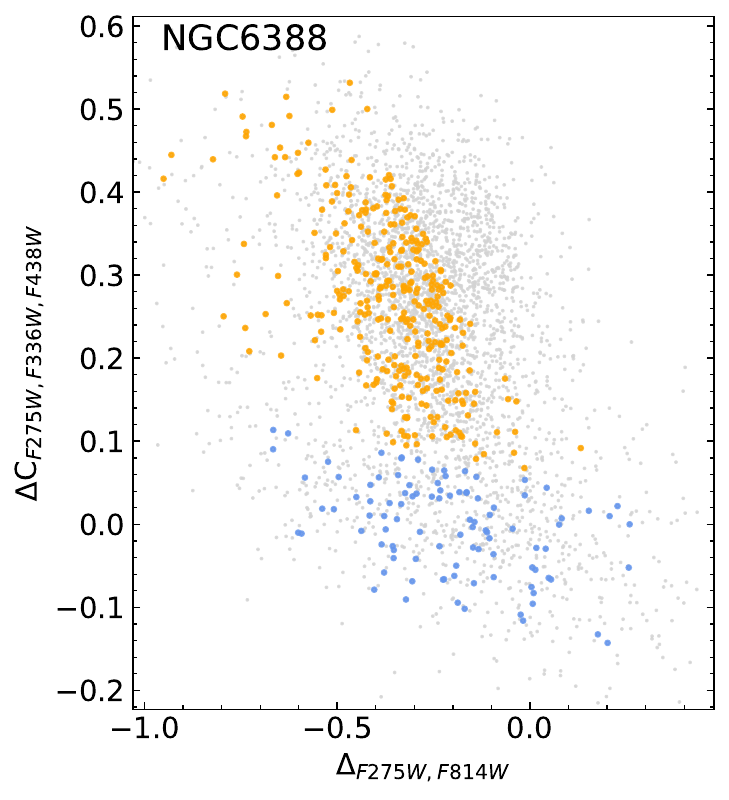}
   }   
     \caption{Same as Fig.~\ref{fig:app_cmap1} but for NGC\,2808, NGC\,3201, NGC\,5286, NGC\,5904, NGC\,6093, NGC\,6218, NGC\,6254, NGC\,6362 and NGC\,6388.}\label{fig:app_cmap2}
\end{figure}

\begin{figure}
\resizebox{\hsize}{!}{   
     \includegraphics{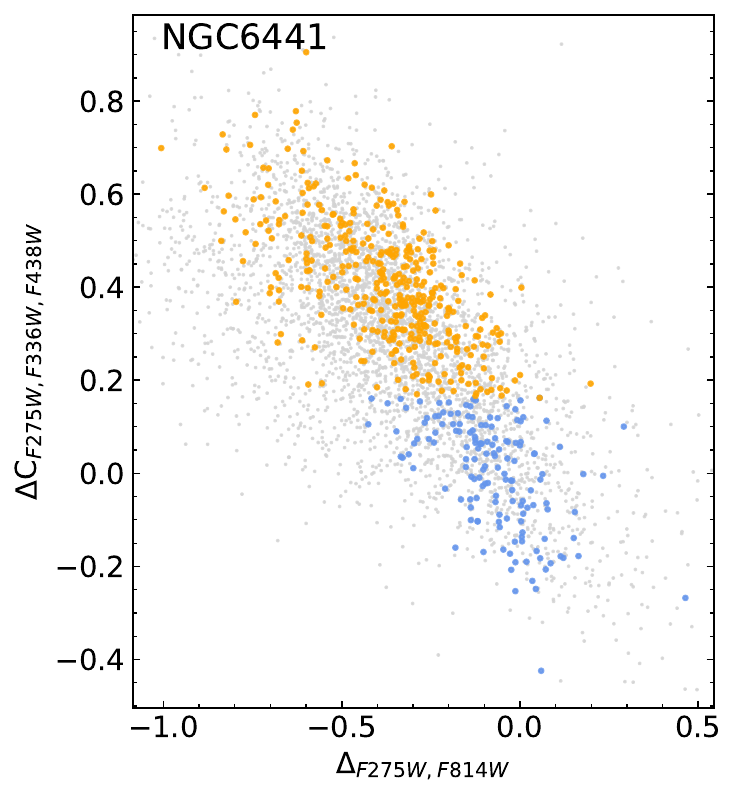}
   \includegraphics{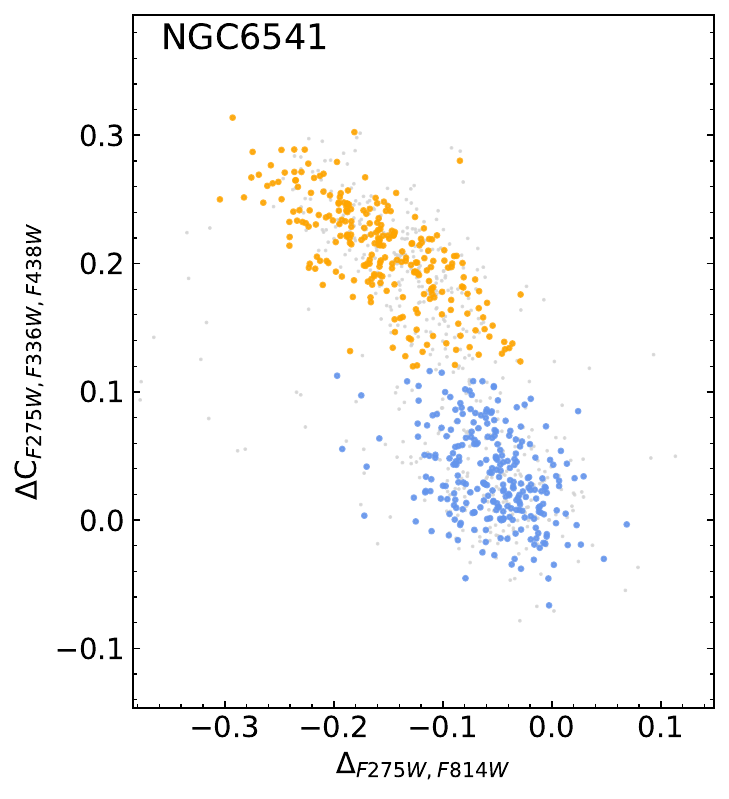}
   \includegraphics{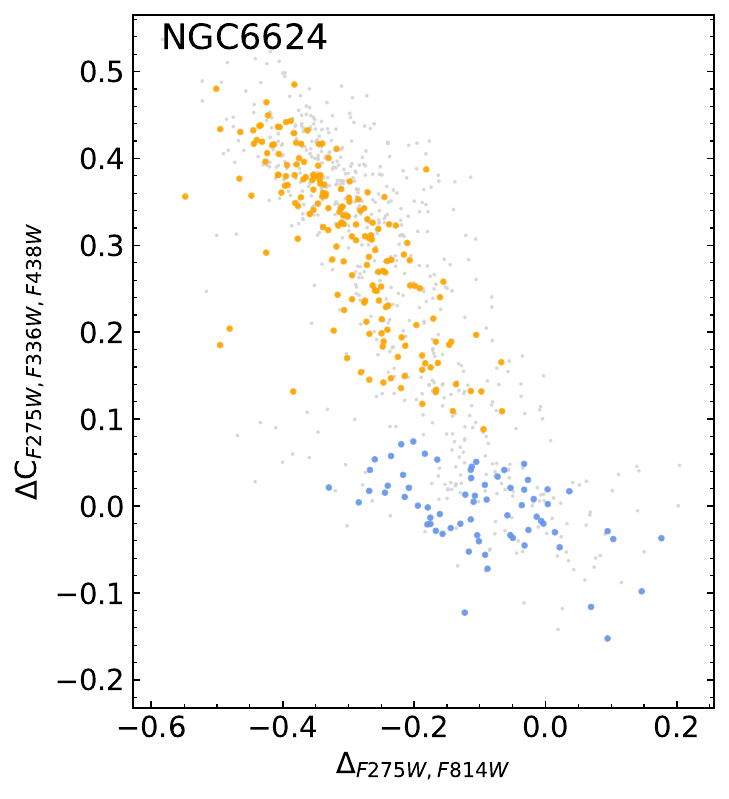}
   }\vspace{1pt}
\resizebox{\hsize}{!}{
   \includegraphics{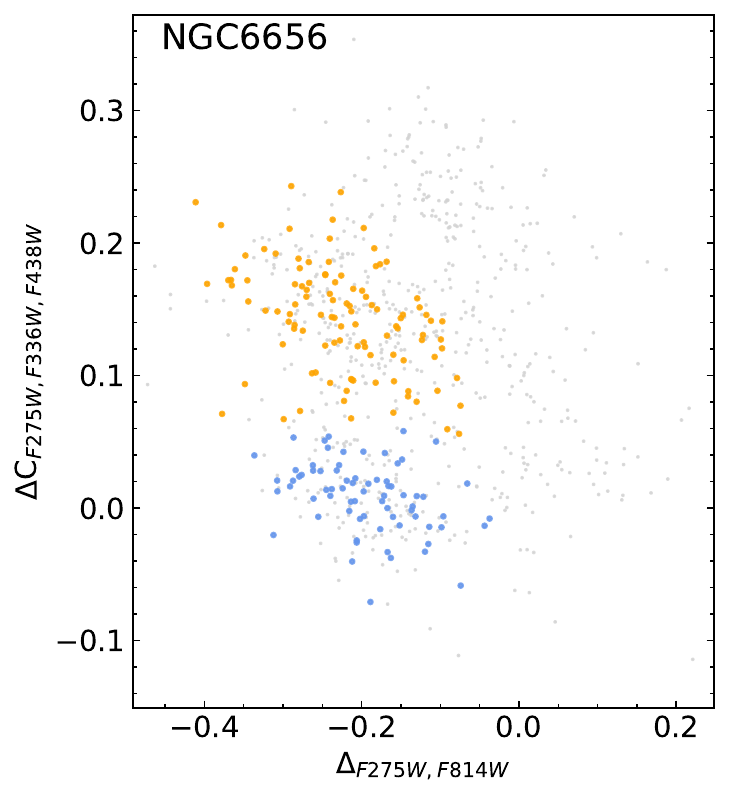}
   \includegraphics{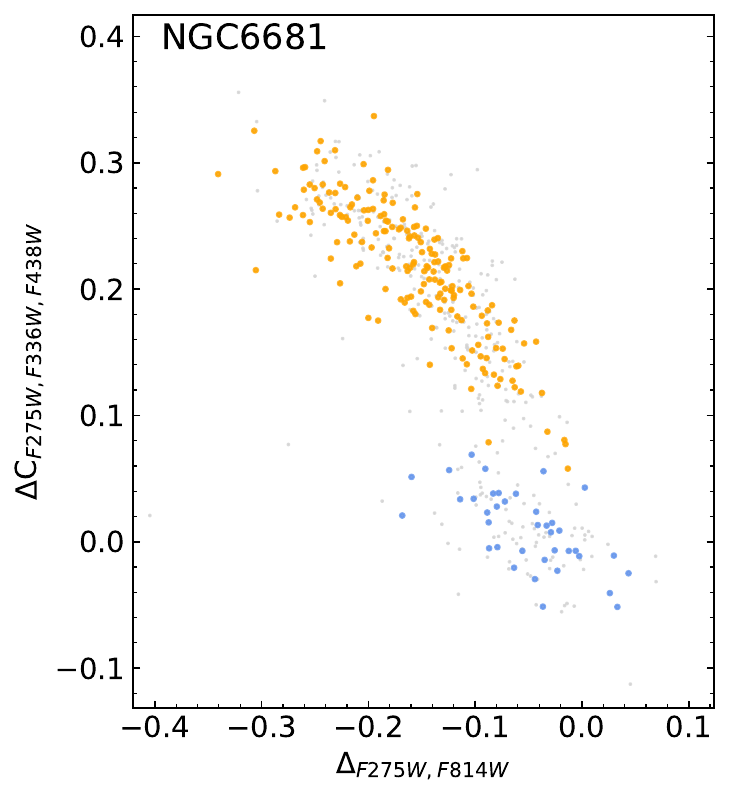}
   \includegraphics{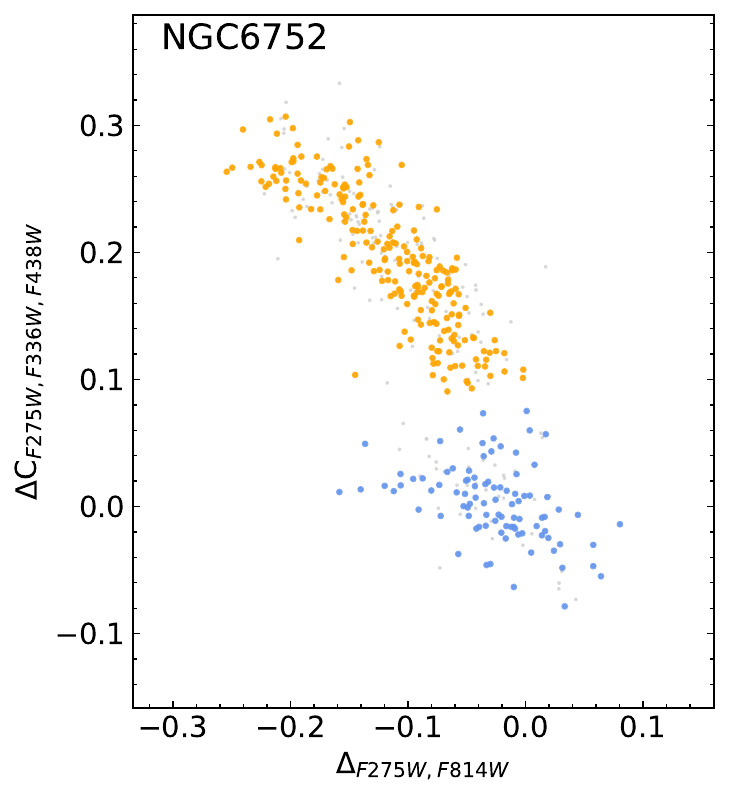}
   }  \vspace{1pt}
   \resizebox{\hsize}{!}{
   \includegraphics{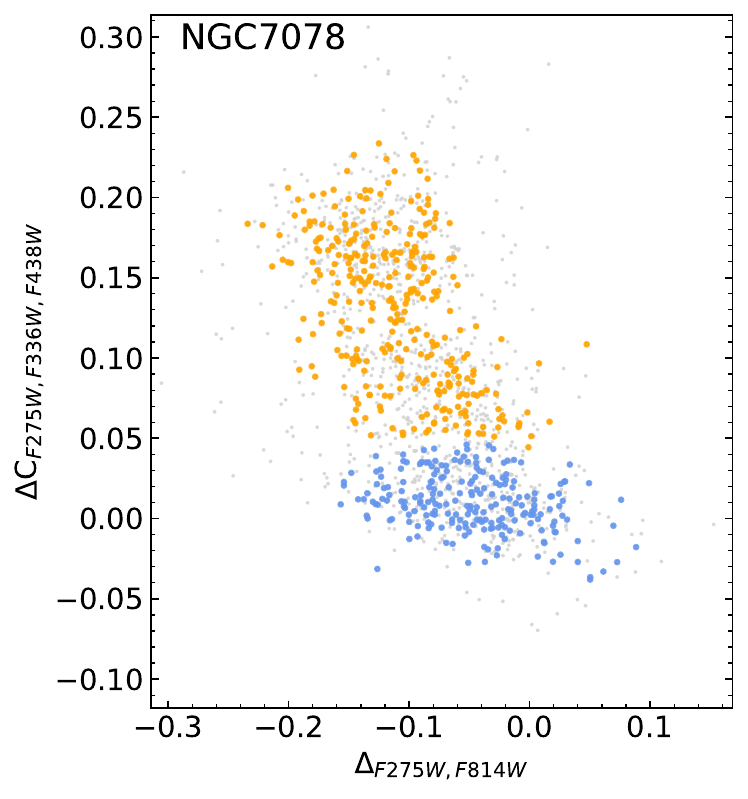}
   \includegraphics{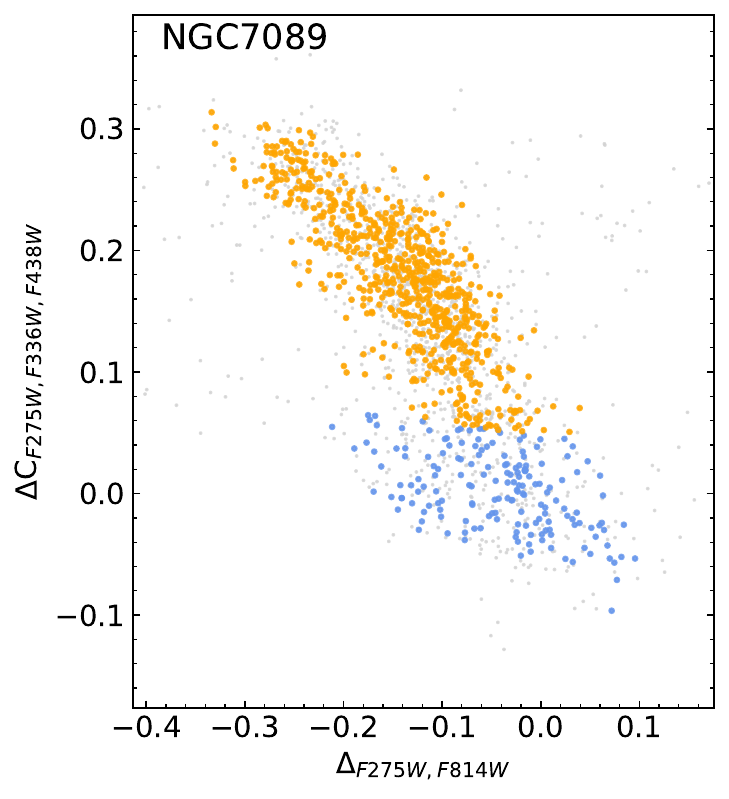}
   \includegraphics{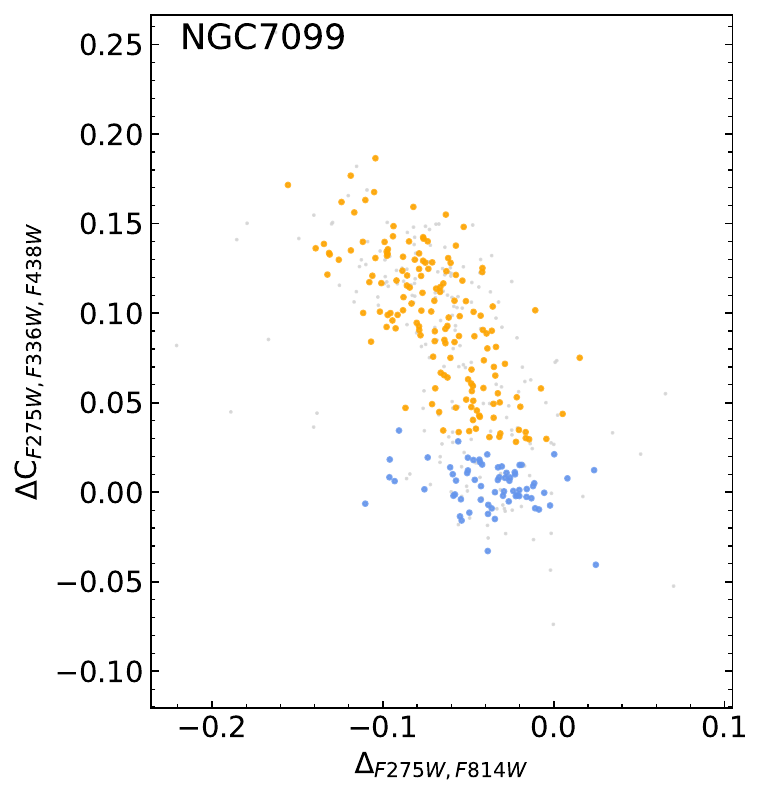}   }
     \caption{Same as Fig.~\ref{fig:app_cmap1} but for NGC\,6441, NGC\,6541, NGC\,6624, NGC\,6656, NGC\,6681, NGC\,6752, NGC\,7078, NGC\,7089, and NGC\,7099.}\label{fig:app_cmap3}
\end{figure}


\end{appendix}

\end{document}